\documentclass{article}
\usepackage[utf8]{inputenc}
\usepackage{xcolor}
\usepackage{amsmath}
\usepackage{amssymb}
\usepackage{authblk} 
\usepackage[left]{lineno}
\usepackage{lineno}
\usepackage{caption}
\usepackage{subcaption}
\usepackage[numbers,sort&compress]{natbib}
\usepackage{graphicx}
\usepackage{doi}
\usepackage{enumerate}
\DeclareUnicodeCharacter{2060}{\nolinebreak}

\usepackage[utf8]{inputenc} % allow utf-8 input
\usepackage[T1]{fontenc}    % use 8-bit T1 fonts
\usepackage{hyperref}       % hyperlinks
\hypersetup{backref=true,       
    pagebackref=true,               
    hyperindex=true,                
    colorlinks=true,                
    breaklinks=true,                
    urlcolor= blue,                
    linkcolor= blue,                
    bookmarks=true,                 
    bookmarksopen=false,
    filecolor=black,
    citecolor=blue,
    linkbordercolor=red}
\AtBeginDocument{\hypersetup{pdfborder={0 0 0}}}   
\usepackage{setspace}
\usepackage{geometry}
\title{\textbf{Theoretical and Experimental Challenges in the Measurement of Neutrino Mass}}
\author[$1$]{Jyotsna Singh} %\thanks{corresponding author: singh.jyotsnalu@gmail.com}}
\author[$2$]{M. Ibrahim Mirza \thanks{corresponding author: ibm.lhcms@gmail.com}}

\affil[$1$]{\textit{\small{Department of Physics, University of Lucknow, Lucknow, Uttar Pradesh, India}}}
\affil[$2$]{\textit{\small{Department of Physics and Astronomy, University of Tennessee, Knoxville, Tennessee 37916, USA}}}
\date{}
\begin{document}
\maketitle
\begin{abstract}
Neutrino masses are yet unknown. We discuss the present state of effective electron anti-neutrino mass from $\beta$ decay experiments; effective Majorana neutrino mass from neutrinoless double-beta decay experiments; neutrino mass squared differences from neutrino oscillation: solar, atmospheric, reactor and accelerator based experiments; sum of neutrino masses from cosmological observations. Current experimental challenges in the determination of neutrino masses are briefly discussed. The main focus is devoted to contemporary experiments. 
\end{abstract}
\section{Introduction} 
Neutrinos are the second most abundant known particles in the Universe. Despite of their abundance in the nature, their hypothetical presence was first announced by Wolfgang Pauli in 1930, when trying to protect the law of conservation of energy in beta radioactivity \cite{pauli}. This particle got its name "Neutrino" by Enrico Fermi in 1934. The neutrinos were introduced as the neutral and massless fermions \cite{REINES:1956ug}. These neutrinos interact only via weak interaction and their cross section of interaction is very small \cite{PhysRev.92.830}. 
\\ The Standard Model (SM) of particle physics is based on gauge group (SU(3)$_{\text{C}}\times$ SU(2)$_{\text{L}}\times$ U(1)$_{\text{Y}}$) \cite{GLASHOW1961579,PhysRevLett.19.1264}. The electroweak group is represented by SU(2)$_{\text{L}}\times$U(1)$_{\text{Y}}$. SM describes the interaction between fundamental matter particles i.e. quarks and leptons which are fermions, three fields i.e. electromagnetic, weak and strong field and their associated gauge bosons along with a scalar Higgs boson. All the charged fermions in the SM are Dirac, leaving neutrinos. Neutrinos are Dirac ($\nu \neq \overline{\nu}$) or Majorana ($\nu = \overline{\nu}$) is yet to be established \cite{Majorana:2008vb,PhysRev.56.1184,Bilenky:2018hbz}.  In SM, neutrinos are considered as massless fermion.
\\ The discovery of neutrino oscillations by neutrino experiments came up with the rejection of the idea of massless neutrino. The neutrino oscillation was confirmed by Super Kamiokande \cite{PhysRevLett.81.1562} and Sudbury Neutrino Observatory \cite{PhysRevLett.89.011301}, this remarkable discovery led to the Nobel Prize in Physics in 2015 \cite{RevModPhys.88.030501,RevModPhys.88.030502}. This discovery was the first experimentally confirmed dent in the SM and it opened the door for physics Beyond Standard Model (BSM). 
\section{Neutrino Mass}
Generally the neutrino mass can be determined using, (i) cosmological data : sets a most stringent bound on the sum of neutrino masses ($\Sigma m_{\nu}$), (ii) beta decay : sets a most stringent bound on effective electron anti-neutrino mass ($m_{\nu_e}$) by observing the kinematics of weak interaction, (iii) neutrinoless double-beta decay : sets a most stringent bound on effective Majorana neutrino mass ($m_{\beta\beta}$) by observing the mono-energetic peak (if observed) at the decay $Q$-value. These approaches are discussed below.
%%%
\subsection{Sum of neutrino masses : cosmological bounds}
Sum of neutrino mass is defined as : $\Sigma$m$_\nu$ = m$_1$+m$_2$+m$_3$, where m$_1$, m$_2$ and m$_3$ are three neutrino mass eigenstates. Cosmological observations carry imprints of neutrinos and therefore it can be used to extract and constrain the neutrino properties. Cosmology is sensitive to the following neutrino properties: (i) number of active neutrinos, (ii) neutrino density, (iii) sum of neutrino masses. 
\\ Generally accepted cosmological model, Standard Model of cosmology explain the large scale structures, their dynamics and to answer unresolved puzzles associated with the evolution and fate of the Universe. The $\Lambda$-CDM (cold dark matter) model best describes the present parameters, such as density parameter of baryons ($\Omega_b$ $\simeq$ 0.05) refers to observable objects in the Universe, density parameter of CDM ($\Omega_c$ $\simeq$ 0.25) refers to non-baryonic and nonrelativistic matter, density parameter of cosmological constant ($\Omega_{\Lambda}$ $\simeq$ 0.70) refers to vacuum, also called the dark energy and Hubble constant (h $\simeq$ 70 km s$^{-1}$ Mpc$^{-1}$) refers to present rate of expansion of the Universe. 
\\ The precise estimation of neutrino to photon number density ratio (n$_\nu$/n$_\gamma$) is important for the determination of sum of neutrino masses ($\Sigma m_{\nu}$) and this ratio is fixed in SM including many extensions of SM. The ratio related to $\Sigma m_{\nu}$ as $\Omega_{\nu}$ = $\rho^0_{\nu}/\rho^0_{\text{crit}}$ = $\Sigma m_{\nu}$/($93.14$ h$^2$eV), where $\Omega_{\nu}$ is the present total neutrino density in terms of critical density $\rho^0_{\text{crit}}$. The expression of $\Omega_{\nu}$ reported in the text assumes the "standard" (n$_\nu$/n$_\gamma$). This ratio is connected to the physics of neutrino decoupling.
\\ The neutrino to photon energy density ratio ($\rho_{\nu}/\rho_{\gamma}$) between the $e^-e^+$ annihilation time and non relativistic transition time of neutrino can be given by the expression $\rho_{\nu}/\rho_{\gamma}$ = $(7/8)$ N$_{\text{eff}}$(4/11)$^{4/3}$, where N$_{\text{eff}}$ is effective number of neutrinos estimated as 3.044 from a detailed calculations of the process of neutrino decoupling with at least 10$^{-4}$ numerical precision \cite{mv16,mv17,kensu,kensu2}. The direct measurement of the invisible width of Z-boson limits the number of active left-handed neutrino states to three, N$_{\nu}$ = 2.9963 $\pm$ 0.0074, they are $\nu_e,\nu_{\mu},\nu_{\tau}$ \cite{ParticleDataGroup:2020ssz}. 
\\ The estimated sum of neutrino masses from composite sample (such as Planck, BAO, RSD) based on $\Lambda$CDM+$\Sigma m_{\nu}$ model is mentioned in the table \ref{table:table1}. There are several challenges in measuring sum of neutrino masses which needs to be mentioned for imposing more stringent constraints on $\Sigma m_{\nu}$. Detailed overview of the cosmological constraints on the neutrino properties can found in \cite{julien,pastor,gerbino,synergy}.
\\ 
\\
\textbf{Main challenges in the measurement of sum of neutrino masses}
\begin{itemize}
    \item Measurement of cosmological parameters with utmost accuracy.
    \item Dependency on cosmological model.
    \item Making scaling to current detectors.
    \item Removal of false B-mode signal in the CMB (cosmic microwave background) measurement. 
    \item Sub-percent level precision in BAO (baryon acoustic oscillation) measurements of the distance scale.
    \item Sum of neutrino mass calculated by different models using composite dataset has to be minimized, because we know, if the neutrino mass variation is in the range 0.025 eV - 1 eV then the error of the order of 5$\%$ will be generated on the matter power spectrum in comparison to the current matter power spectrum.
\end{itemize}
Next generation cosmological experiments will addresses above mentioned issues and provide better constraints on $\Sigma m_{\nu}$, few upcoming experiment are, DESI \cite{mnj}, Euclid \cite{bhj}, LSST \cite{azxc}, SPHEREx \cite{klj}, SKA \cite{mnhu}, Simon Observatory \cite{sde}, CMB-S4 \cite{fgfs}, LiteBird \cite{dfw}.
\\
\begin{table}[h!]
\centering
\begin{tabular}{lc lc lc}
\\
\hline
\hline
\textbf{Data} & \textbf{$\Sigma$}$\textbf{m}_{\nu}$ \scriptsize{(eV)} & \textbf{Ref.}  \\  %
&\scriptsize{(95\% C.L.)}&\\
\hline
\hline
Planck (TT+low-E) & < 0.54 & \cite{cosmu1} \\
Planck (TT,TE,EE+low-E) & < 0.26 & \cite{cosmu1} \\
Planck (TT+low-E)+BAO & < 0.13 & \cite{des}\\
Planck (TT+low-E+lensing) & < 0.44 & \cite{cosmu1} \\
Planck (TT,TE,EE+low-E+lensing) & < 0.24 & \cite{cosmu1} \\
Planck (TT,TE,EE+low-E)+BAO+RSD & < 0.10 & \cite{des}\\
Planck (TT+low-E+lensing)+BAO+Lyman-$\alpha$ & < 0.087 & \cite{hsa}\\
Planck (TT,TE,EE+low-E)+BAO+RSD+Pantheon+DES & < 0.13 & \cite{dsds}\\
\hline
\hline
\end{tabular}
\caption{Cosmological bounds on sum of neutrino masses based on $\Lambda$CDM+$\Sigma m_{\nu}$ model. Where, TT: temperature power spectra, TE: temperature - polarization power spectra, EE: polarization power spectra, low-E: low-$l$ polarization, RSD: Redshift-Space Distortions, DES: Dark Energy Survey, Pantheon: combined sample of supernova Type Ia.}
\label{table:table1}
\end{table}
\subsection{Effective electron anti-neutrino mass : $\beta$ decay bounds}
\begin{figure}[h!]
    \includegraphics[width=7.7cm]{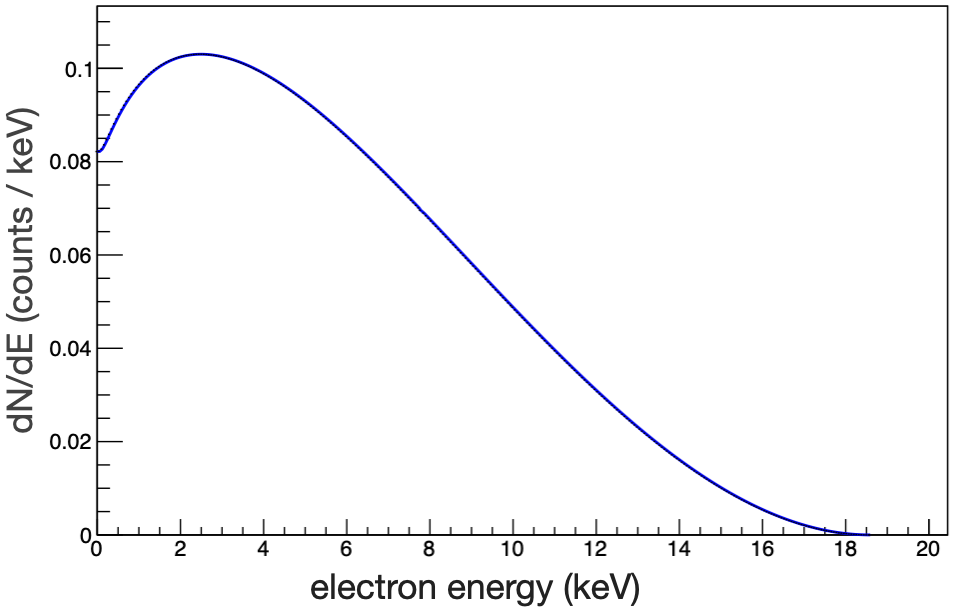}
    \hspace{0.1cm}
    \includegraphics[width=8cm]{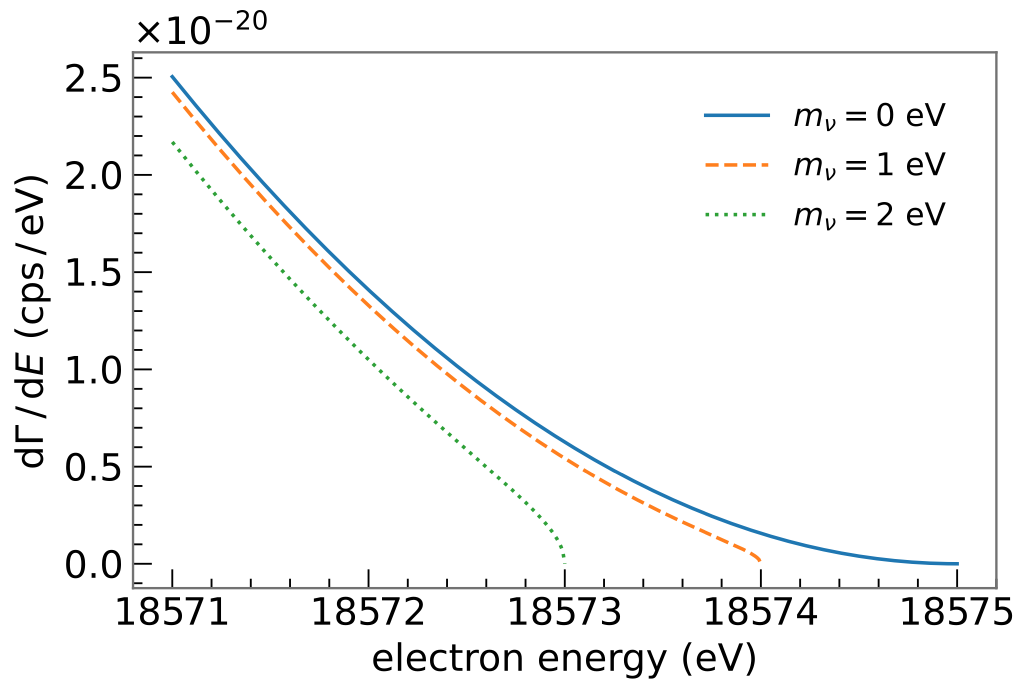}
    \caption{\textbf{Left} : The beta spectrum of ${}^{3}$H. Data is taken from \cite{nndc}. \textbf{Right} : Shape distortion produced by effective electron anti-neutrino mass (m$_\nu$) in ${}^{3}$H beta spectrum. Image credit \cite{sh3h}.}
     \label{fig:3hfig}
\end{figure}

Effective electron anti-neutrino mass is defined as : m$_{\nu_e}$ $\equiv$ $\sqrt{|\text{U}_{e1}|^2\text{m}_1^2+|\text{U}_{e2}|^2\text{m}_2^2+|\text{U}_{e3}|^2\text{m}_3^2}$, where U$_{e1}$, U$_{e2}$, and U$_{e3}$ are components of neutrino mixing matrix. Determination of m$_{\nu_e}$ is of urgent importance for cosmology, particle physics. This information will help in understanding the role of neutrinos in the structure formation of the Universe after the Big Bang \cite{aqqw}. In addition, the value of effective electron anti-neutrino mass will help us in identifying the right theories for the prediction of BSM physics \cite{cxd,ghg}.
\clearpage
The $\beta$ decay experiments are designed in a way to explore effective electron anti-neutrino mass. The weak interaction process of $\beta$ decay can be expressed as n$\rightarrow$ p+e$^-$+$\overline{\nu_e}$, a careful study of the given reaction can quantitate the effective electron anti-neutrino mass. $\beta$-decay experiment measures a distortion in the spectral shape near the endpoint of $\beta$ decaying isotope. The phase space of an electron emitted in a $\beta$ decay process can be expressed as P(E) $\propto E(E-E_0) p \sqrt{(E_0-E)^2-m_{\nu_e}^2}$ where $p$ is the momentum of outgoing electron possessing energy E, $m_{\nu_e}$ is the effective electron anti-neutrino mass and $E_0$ is end point energy of the spectrum. The $\beta$ decay spectrum of ${}^{3}$H (Q$_{\beta}$=18.6 keV) is shown in figure \ref{fig:3hfig} (left) and distortion produced by effective electron anti-neutrino mass in ${}^{3}$H energy spectrum is shown in figure \ref{fig:3hfig} (right).
\\ One of the most promising experiment designed to probe the effective electron anti-neutrino mass by studying the kinematics of $\beta$ decay is KATRIN (Karlsruhe Tritium Neutrino) experiment by looking at the decay of tritium (${}^{3}$H) as ${}^{3}$H $\rightarrow$ ${}^{3}$He + e$^-$+$\overline{\nu_e}$. KATRIN put a constrain on $m_{\nu_e}<$ 0.8 eV at the 90$\%$ C.L \cite{katrin} and have potential to impose constrain on $m_{\nu_e}<$ 0.2 eV.
\\ KATRIN reduce the statistical uncertainty by a factor of three and systematic uncertainty by a factor of two relative to its earlier campaign. In a first campaign KATRIN (2019) reached a sensitivity of 1.1 eV at 90$\%$ C.L and in its second campaign KATRIN (2021) achieved sensitivity of 0.7 eV at 90$\%$ C.L. KATRIN would be dominated by systematics, although results of KATRIN first and second campaign are dominated by statistical uncertainties. KATRIN continue reducing its systematic uncertainty to achieve a designed sensitivity of 0.2 eV at 90$\%$ C.L on $m_{\nu_e}$. The constrain imposed on upper limit of effective electron anti-neutrino mass by KATRIN experiment is shown in table $\ref{table:table2}$. To pin down the effective electron anti-neutrino mass from $\beta$ decay experiments many potential challenges needs to be addressed carefully. 
\\ 
\\
\textbf{Main challenges in the measurement of effective electron anti-neutrino mass}
\begin{itemize}
    \item Spectrometer with good counting rate or high efficiency.
    \item Spectrometer with better end-point energy resolution.
    \item Intense source of tritium and Holmium-163 : high Becquerel activity is recommended.
    \item Energy loss of $\beta$ in the source, \cite{Bilenky:2018hbz,beta8}.
    \item Removal of background produced by radon decays inside spectrometer.
\end{itemize}
Many other promising upcoming experiments which are designed to impose better constrain on $m_{\nu_e}$ are PTOLEMY (${}^{3}$H) \cite{Betti_2019}, Project8 (${}^{3}$H) \cite{PhysRevD.80.051301}, EcHo (${}^{163}$Ho) \cite{Gastaldo:2014un}, HOLMES (${}^{163}$Ho) \cite{Alpert:2015vi}, NuMECS (${}^{163}$Ho) \cite{Croce:2016uq}.
\\ The advantage of using ${}^{163}$Ho isotope having 100$\%$ decay via electron capture process and very small total nuclear decay energy ($<$ 3 keV).
\begin{table}[h!]
\centering
\begin{tabular}{lc c c}
\\
\hline
\hline
\textbf{Experiments} & \textbf{Isotope} & \textbf{m}$_{\nu_e}$ (eV) &\textbf{Ref.}\\ [0.5ex] %
%%%
&&\scriptsize{(90\% C.L.)}&\\[0.5ex]
\hline
\hline
KATRIN (2019)& ${}^{3}$H &  < 1.1  &\cite{kat1}\\
KATRIN (2021)& ${}^{3}$H &  < 0.9 &\cite{katrin}\\
KATRIN (combined)& ${}^{3}$H &  < 0.8  &\cite{katrin}\\
\hline
\hline
\end{tabular}
\caption{Current bounds on effective electron anti-neutrino mass from $\beta$ decay kinematics.}
\label{table:table2}
\end{table}
\subsection{Effective Majorana neutrino mass : 0$\nu\beta\beta$ decay bounds}
\begin{figure}[h!]
    \includegraphics[width=8cm,height=5cm]{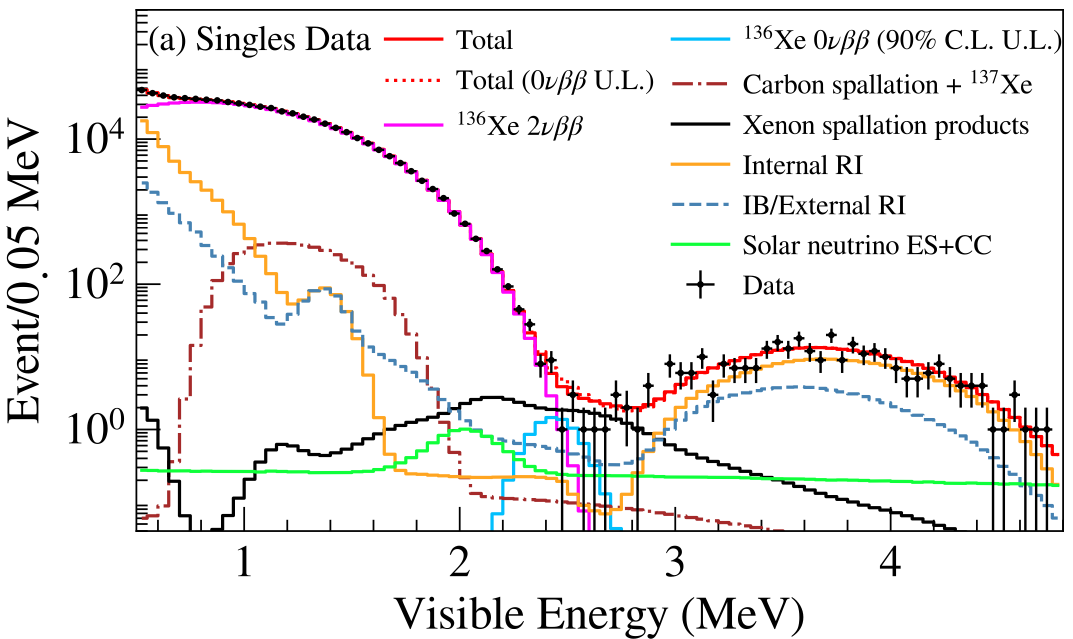}
    \hspace{0.1cm}
    \includegraphics[width=8cm,height=5cm]{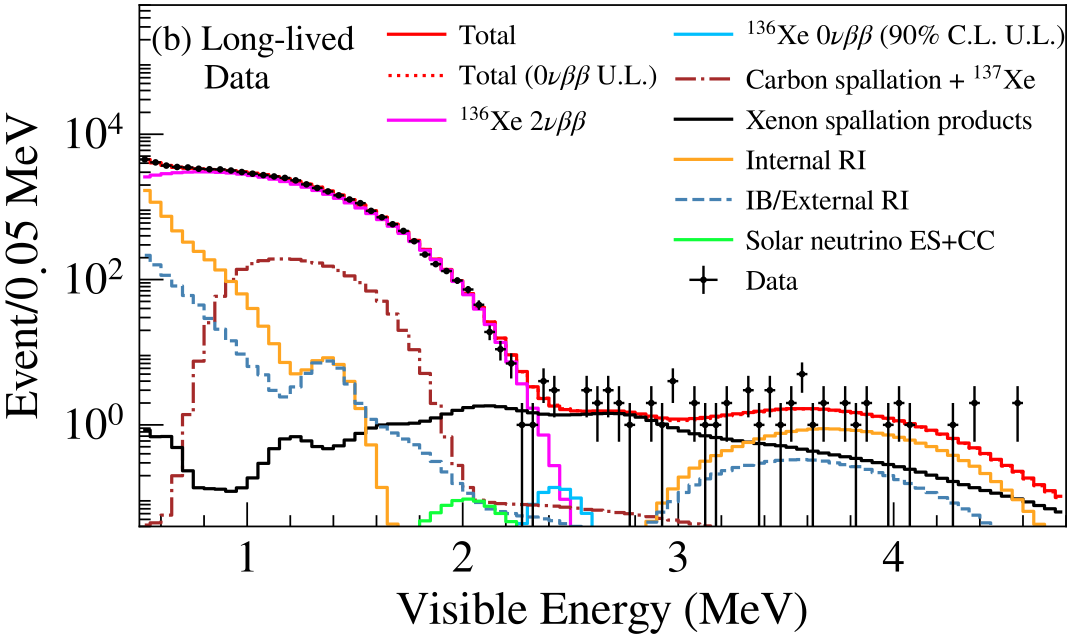}
    \caption{Energy distribution measured by KamLAND-Zen experiment. \textbf{Left} : events from short-lived backgrounds. \textbf{Right} : events from long-lived backgrounds. Fit to the 0$\nu\beta\beta$ decay signal Q$_{\beta\beta}$ at 2.458 MeV is shown in cyan. Image credit \cite{zen}.} 
     \label{fig:kamla}
\end{figure}

\begin{figure}[h!]
    \includegraphics[width=\textwidth]{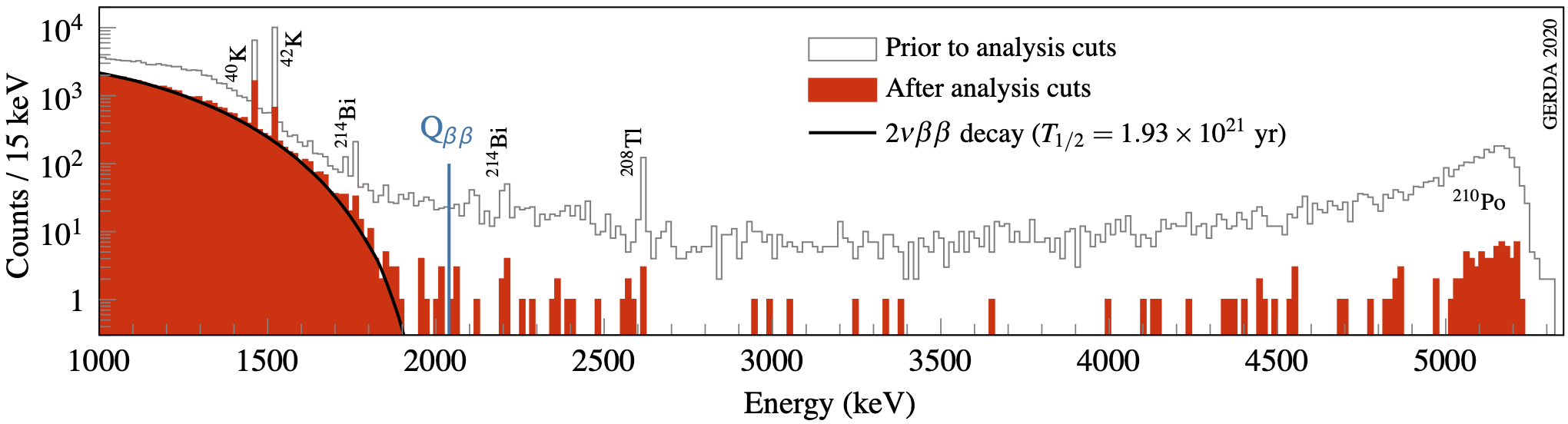}
    \caption{Energy distributed measured by GERDA experiment, expected 0$\nu\beta\beta$ decay peak Q$_{\beta\beta}$ at 2039 keV is shown in blue. Image credit \cite{PhysRevLett.125.252502}.} 
     \label{fig:gerd}
\end{figure}

Effective Majorana neutrino mass is defined as : m$_{\beta\beta}$ $\equiv$ |U$_{e1}^2$m$_1$+U$_{e2}^2$m$_2$+U$_{e3}^2$m$_3$|. Neutrinoless double-beta (0$\nu\beta\beta$) decay is a hypothetical nuclear transition and is expressed as (A,Z) $\rightarrow$ (A,Z+2) + 2$e^-$. This lepton number violating \cite{valle} phenomenon if observed, will assign neutrinos Majorana characteristics of particle. 0$\nu\beta\beta$ decay could provides effective Majorana neutrino mass assuming the decay is mediated by light Majorana neutrino. Experiments measuring 0$\nu\beta\beta$ decay measure small peak generated by sum of energy of energy of two electrons.
%%%
\\ Different isotopes used by experiments searching for the signatures of 0$\nu\beta\beta$ decay are, ${}^{76}$Ge \cite{sci,PhysRevLett.125.252502}, ${}^{82}$Se \cite{bn}, ${}^{100}$Mo \cite{PhysRevLett.126.181802}, ${}^{130}$Te \cite{bgh}, ${}^{136}$Xe \cite{PhysRevLett.117.082503,PhysRevLett.123.161802}, ${}^{48}$Ca \cite{heer}, ${}^{96}$Zr \cite{ssdf}, ${}^{116}$Cd \cite{serr}, ${}^{150}$Nd \cite{fdf}. 
The half-life sensitivity of an experiment estimated using the expression,  T$^{0\nu}_{1/2}\propto a \varepsilon \sqrt{\mathcal{E}/(B.\Delta E)}$ (with background) and T$^{0\nu}_{1/2}\propto a \varepsilon \mathcal{E}$ (background free), where a is the isotopic abundance, $\varepsilon$ is the efficiency of the detection of 0$\nu\beta\beta$ signal at the region of interest (ROI), B is the background index, $\Delta E$ is the energy resolution of the detector and the exposure ($\mathcal{E}$) is given by the product of mass of the isotope and run time. 
\\ KamLAND-Zen sets a strongest limit on half-life of any 0$\nu\beta\beta$ decay isotope to date, for ${}^{136}$Xe as T$^{0\nu}_{1/2}$ > 2.3 $\times$ 10$^{26}$ yr \cite{zen}. Energy spectrum of ${}^{136}$Xe measured by KamLAND-Zen experiment (currently running) is shown figure \ref{fig:kamla}. Energy spectrum of ${}^{76}$Ge measured by GERDA experiment (final results) is shown figure \ref{fig:gerd}.
\\
\\  By considering neutrinos to be a Majorana particle its $m_{\beta\beta}$ is estimated for experimentally measured isotopic half-life using relation, (T$^{0\nu}_{1/2})^{-1}$ = G$_{0\nu}$ g$_A^4$ |M$_{0\nu}|^2$ (m$^2_{\beta\beta}/m^2_e$). Here M$_{0\nu}$ is the nuclear matrix element, G$_{0\nu}$ is the phase space factor, g$_A$ is axial coupling constant, m$_e$ is mass of electron. Tightest bounds imposed on the T$^{0\nu}_{1/2}$ of different isotopes by various experiment and estimated $m_{\beta\beta}$ values are mentioned in table \ref{table:table3}.
\\
%%%
\clearpage
\textbf{Main challenges in the measurement of effective Majorana neutrino mass}
\begin{itemize}
    \item Enhanced energy resolution of detectors, to distinctly visualize the mono-energetic signal from the prominent two neutrino double-beta (2$\nu\beta\beta$) decay continuum and to reduce the background since sharper signal sits on less background.
    \item Mitigation of background, it is extremely challenging, experiments are investigating various techniques to minimize the background at the ROI, \cite{chamber,denys,mirza}.
    \item Requirement of large mass of the enriched $\beta\beta$ isotope, to enhance the statistics.
    \item Uncertainty in nuclear matrix element (model-dependent) leads to the uncertainty in m$_{\beta\beta}$ estimation.
\end{itemize}
 LEGEND-200 (${}^{76}$Ge) \cite{LEGEND:2021bnm} experiment is currently taking data. Next generation experiments are planned to address the some of the above challenges. Few upcoming experiments are; AMoRE-II (${}^{100}$Mo) \cite{amore}, nEXO (${}^{136}$Xe) \cite{Adhikari_2021}, SNO+ (${}^{130}$Te) \cite{sno}, SuperNEMO  (${}^{82}$Se) \cite{supernemo}, LEGEND-1000 (${}^{76}$Ge) \cite{LEGEND:2021bnm}, KamLAND-Zen 800 (${}^{136}$Xe) \cite{kam}, NEXT-100  (${}^{136}$Xe) \cite{next}, (isotopes corresponding to each experiments shown in bracket).
\\
\begin{table}[h!]
\centering
\begin{tabular}{lc lc lc lc}
\\
\hline
\hline
\textbf{Experiment} & \textbf{Isotope} & \textbf{T$^{0\nu}_{1/2}$} & \textbf{m}$_{\beta\beta}$ &\textbf{Ref.} \\ [0.5ex] % inserts table %heading
&&\scriptsize{(10$^{26}$ yr)}& \scriptsize{(eV)}&\\
\hline
\hline
KamLAND-Zen & ${}^{136}$Xe & $>$ 2.3 & $<$ 0.036 - 0.156 & \cite{zen} \\
GERDA & ${}^{76}$Ge & $>$ 1.8 & $<$ 0.08 - 0.18 & \cite{PhysRevLett.125.252502} \\
Majorana Demonstrator & ${}^{76}$Ge & $>$ 0.83 & $<$ 0.113 - 0.269 & \cite{mjd23} \\
EXO-200 & ${}^{136}$Xe & $>$ 0.35 & $<$ 0.09 - 0.29 & \cite{Adhikari_2021} \\
CUORE & ${}^{130}$Te & $>$ 0.22 & $<$ 0.09 - 0.31 & \cite{bgh}\\
CUPID-0 & ${}^{82}$Se & $>$ 0.046 & $<$ 0.263 - 0.545 & \cite{bn}\\
CUPID-Mo & ${}^{100}$Mo & $>$ 0.015 & $<$ 0.31 - 0.54 & \cite{PhysRevLett.126.181802}\\[1ex]
\hline
\hline
\end{tabular}
\caption{Current bounds on $0\nu\beta\beta$ decay half-life and effective Majorana neutrino mass.}
\label{table:table3}
\end{table}
\subsection{Neutrino mass squared differences}
In the old theory of electroweak interactions, formulated by Glashow, Weinberg and Salam, lepton flavor was conserved and neutrinos were assumed as massless fermions. This simply means that leptons produced in a particular flavor state will remain in that state forever. 
\\ As soon the theory of two component neutrino was developed, B. Pontecorvo proposed the idea of neutrino oscillation in 1957-1958 \cite{1957,1958}. Later, neutrino oscillation or conversion of the neutrino flavor was observed in the solar  \cite{PhysRevLett.89.011301} and atmospheric \cite{PhysRevLett.81.1562} neutrino experiments. Therefore, solar and atmospheric neutrino anomaly was resolved by assigning oscillation phenomenon to neutrino. In 2015, Nobel prize in physics was awarded to T. Kajita and A. B. McDonald for their landmark discovery of neutrino oscillation. These results of neutrino oscillation were subsequently confirmed by reactor experiment, such as, KamLAND \cite{kams,gd} and long baseline experiment, such as, NOvA \cite{nova}. Neutrino oscillation experiment measure appearance or disappearance channel. 
\\
Due to quantum mechanical nature of neutrino, during propagation, neutrino are represented as superposition of three mass eigenstates |$\nu_i$>, (i=1,2,3) and detected as neutrino flavor state |$\nu_{\alpha}$>, ($\alpha$= e, $\mu, \tau$) are related as |$\nu_{\alpha}$> = $\displaystyle\sum _{i=1} ^3$ U$_{\alpha i}$ |$\nu_i$>, where U$_{\alpha i}$ is the mixing matrix or Pontecorvo–Maki–Nakagawa–Sakata (PMNS) matrix, \cite{Bilenky:1978nj,Bilenky:1976wv,Bilenky:2018hbz}.
The neutrino oscillations can be analytically expressed using PMNS matrix and two mass-squared differences of active neutrino, this makes minimum six parameters, solar mixing angle $\theta_{12}$, atmospheric mixing angle $\theta_{23}$, reactor mixing angle $\theta_{13}$, solar mass-squared difference $\Delta m^2_{21}$, atmospheric mass-squared difference $\Delta m^2_{31}$, Dirac CP violating Phase $\delta_{CP}$. In a PMNS matrix $\delta_{CP}$ informs about the difference in neutrino and antineutrino oscillations. For baseline (L), neutrino energy (E) and $\delta_{CP}$ = 0, in three flavor neutrino oscillation probability equations can be expressed as: 
\\
\\ For small values of L/E :
\\ ${}$ \hspace{4cm} P($\nu_e \rightarrow \nu_{\mu}$) = sin$^2(2\theta_{13}$) sin$^2(\theta_{23}$) sin$^2$(1.27$\Delta$m$^2_{23}$$\frac{L}{E}$) 
\\ For large values of L/E :
\\ ${}$ \hspace{4cm} P($\nu_e \rightarrow \nu_{\mu,\tau}$) = cos$^2(\theta_{13}$) sin$^2(2\theta_{12}$) sin$^2$(1.27$\Delta$m$^2_{12}$$\frac{L}{E}$) + $\frac{1}{2}$sin$^2$2$\theta_{13}$
\\
\\ Among the above mentioned six parameters, $\theta_{12},\theta_{13},\theta_{23}$, $\Delta$m$^2_{21}$, and $|\Delta$m$^2_{31}|$ are known with good precision but sign of $|\Delta$m$^2_{31}|$, value of $\delta_{CP}$ and octant of $\theta_{23}$ are still in research phase. Sign of $|\Delta$m$^2_{31}|$ is unknown therefore the mass of active neutrinos can be represented by two hierarchies; normal hierarchy ($\Delta$m$^2_{31}$ > 0) and inverted hierarchy ($\Delta$m$^2_{31}$ < 0). Where mass state distribution in these hierarchy indicates, normal hierarchy : m$_3$ > m$_2$ > m$_1$ and inverted hierarchy: m$_2$ > m$_1$ > m$_3$. Several groups working on global fits to neutrino oscillation data \cite{glodata,sep9,sep90}.
\subsubsection{Solar neutrino experiment}
\begin{figure}[h!]
    \includegraphics[width=8cm]{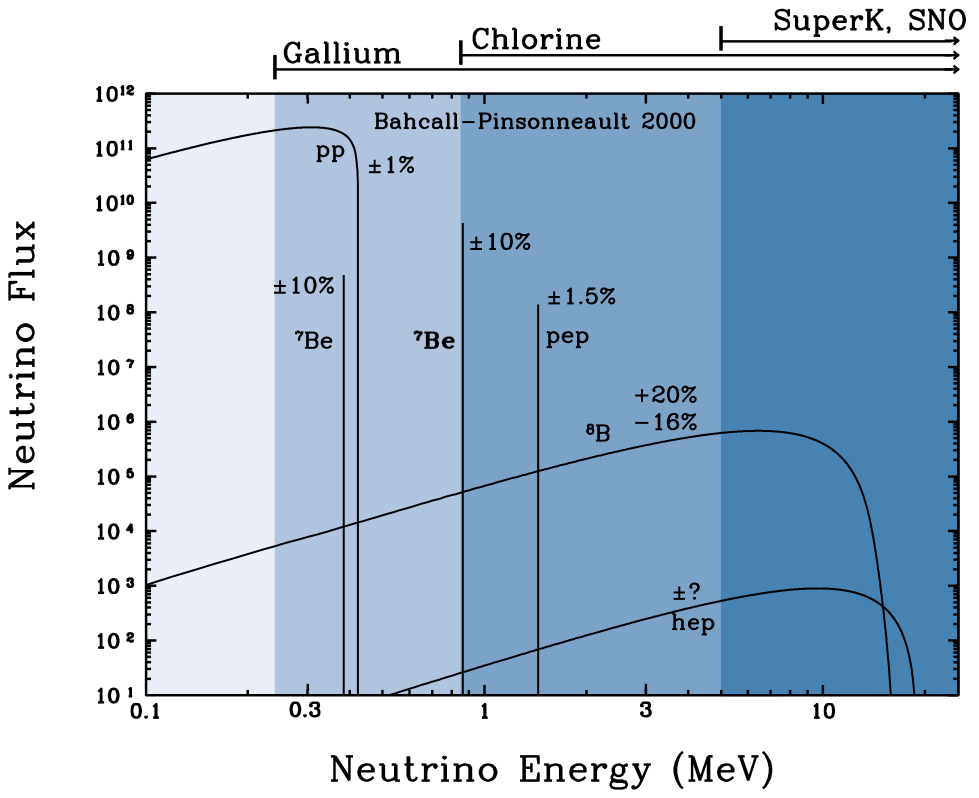}
    \hspace{0.1cm}
    \includegraphics[width=8.5cm]{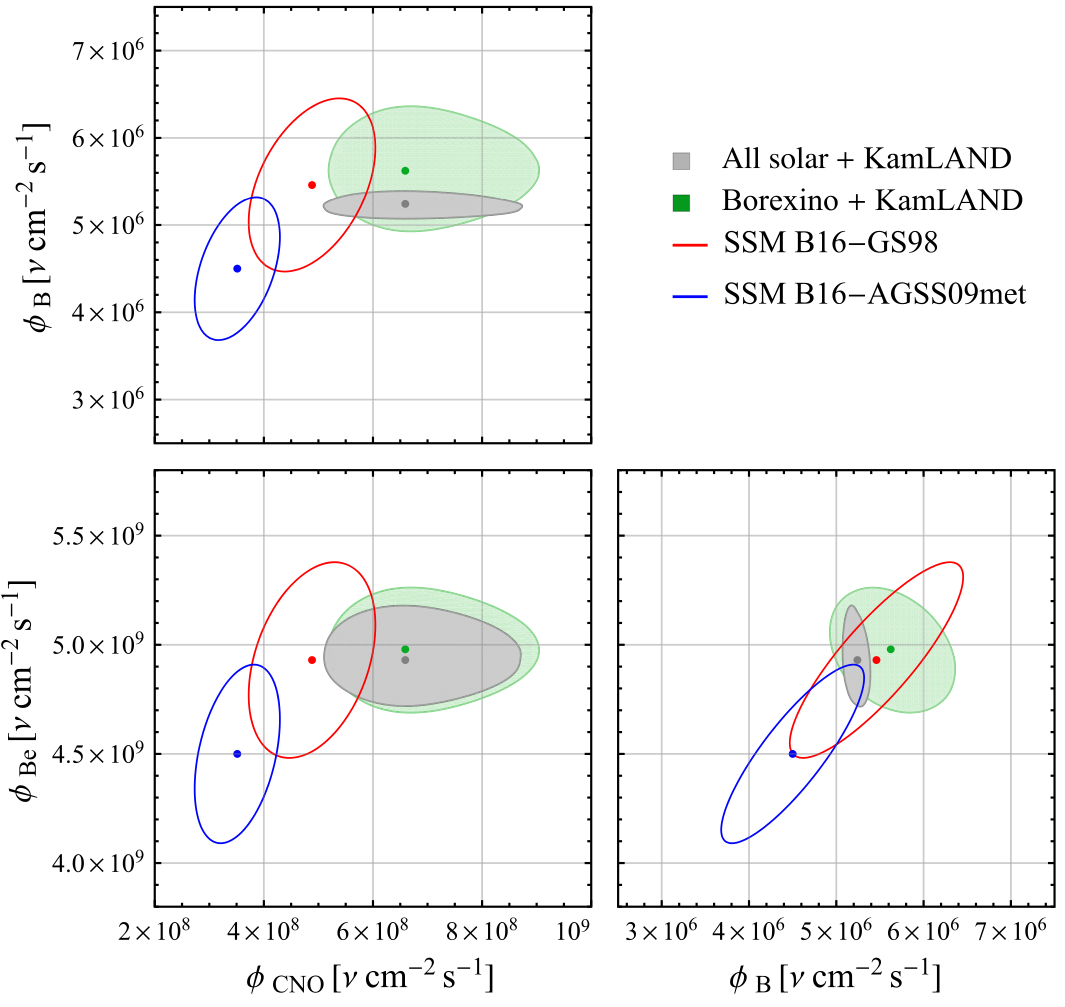}
    \caption{\textbf{Left} : Predicted neutrino flux from Solar Standard Model, image credit \cite{bchall1,bchall2}.
    \textbf{Right} : Estimated 1$\sigma$ allowed region of solar neutrino flux by experiments and SSM, image credit \cite{bors2}.}
     \label{fig:borf}
\end{figure}

\begin{figure}[h!]
    \centering
    \includegraphics[width=9cm,height=8cm]{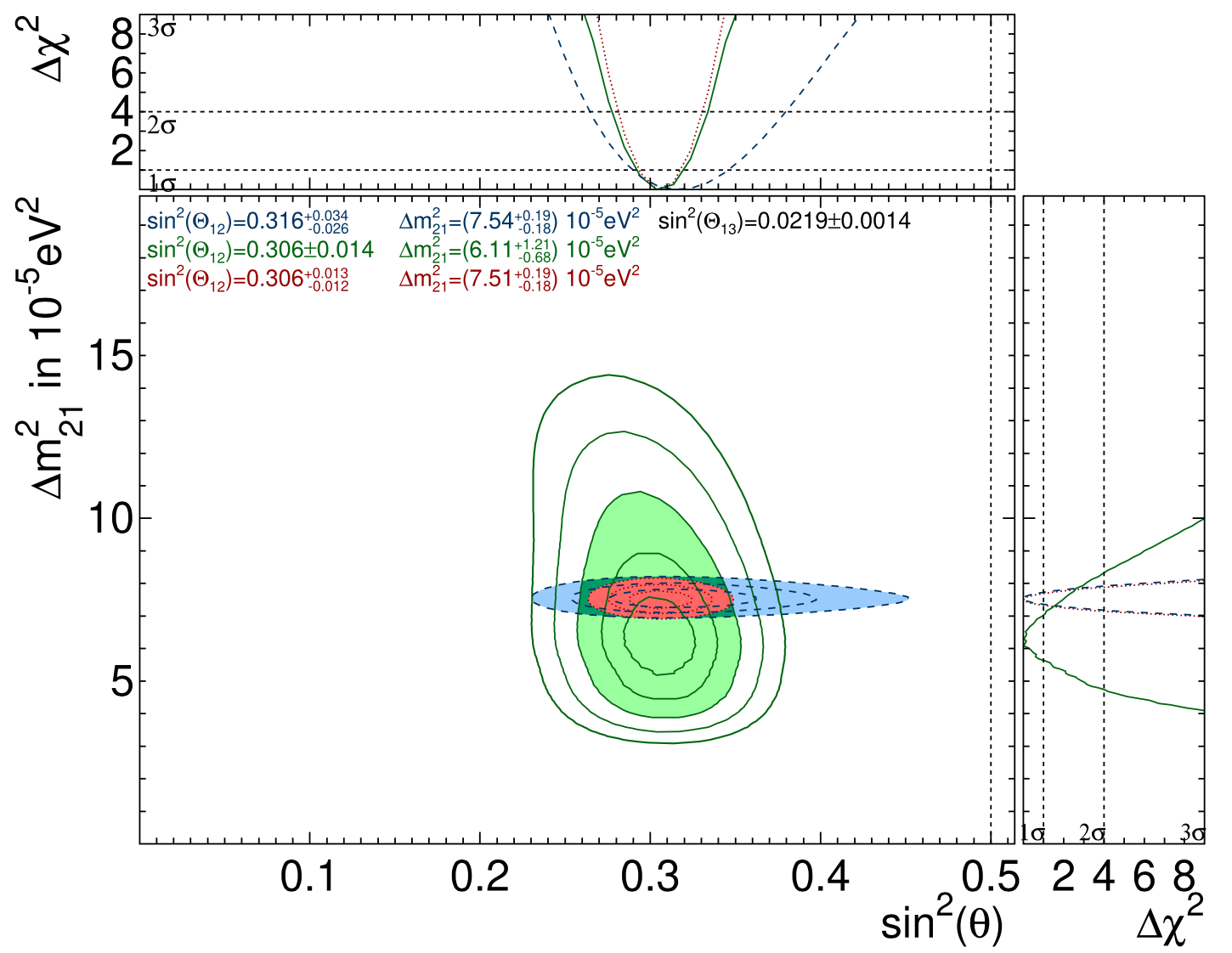}
    \caption{Dependency of $\Delta$m$^2_{21}$ on sin$^2$$\theta_{12}$ from all solar neutrino data for allowed 1$\sigma$, 2$\sigma$, 3$\sigma$ regions where green represents: SK+SNO, blue represents: KamLAND, red represents for combined result. Image credit \cite{neu,nu20}.}
    \label{fig:dm21}
\end{figure}

Sun is a abundant source of neutrino and produce electron neutrino in the process of fusion. The total solar neutrino flux comes from different fusion reactions as shown in \ref{fig:borf} left. Amongst these the dominant contribution to solar neutrino flux comes from pp reaction (99.6\%). 
\\ Solar neutrino experiments designed to study solar neutrinos flux can be largely divided into two groups, (i) radiochemical : (a) gallium based experiment (GALLEX-GNO; SAGE), (b) chlorine based experiment (Homestake); (ii) real time : (a) (heavy) water detectors (Kamiokande; Super-Kamiokande; SNO), (b) liquid scintillator detectors (Borexino; KamLAND). 
\\ The solar neutrino fluxes predicted by Standard Solar Model (SSM) at one astronomical unit is shown in figure \ref{fig:borf} left, where continuum sources are in units of cm$^{-2}$ s$^{-1}$ MeV$^{-1}$, and the line fluxes are in units of cm$^{-2}$ s$^{-1}$. 
\\ Along with the SSM predictions, figure \ref{fig:borf} right, gives the current picture of experimentally estimated flux of B, Be with respect to CNO (carbon–nitrogen–oxygen) and Be with respect to B. These estimated solar neutrino fluxes are compared with solar models, SSM B16-GS98 \cite{ssmb16} and SSM B16-AGSS09met \cite{ssmb16}.
\\ Allowed 1$\sigma$, 2$\sigma$, 3$\sigma$ regions for $\Delta$m$^2_{21}$ and sin$^2$$\theta_{12}$ from all solar neutrino data (green: SK+SNO), (blue: KamLAND), combined result (red) as shown in figure \ref{fig:dm21}. From accelerator and short-baseline reactor neutrino experiments, a combined three-flavor analysis of solar and KamLAND data gives fit values for the oscillation parameter, $\Delta$m$^2_{12}$ = (7.53 $\pm$ 0.18) $\times$ 10$^{-5}$ eV$^2$ \cite{ParticleDataGroup:2020ssz,kama}. Solar neutrino experiments are listed in table \ref{table:table4}.
%%%
\\ 
\textbf{Main challenges in the measurement of mass squared difference from solar neutrinos}
\begin{itemize}
    \item Good energy resolution of detectors.
    \item Large fiducial mass of detector to reduce the statistical error.
    \item Suppression of U, Th, ${}^{14}$C and radon-daughter contamination inside the detector volume.
    \item Detector needs to be shielded by overburden of the Earth to reduce the cosmic background.
    \item Accurate measurement of pp, pep, $^7$Be fluxes in constraining $\Delta$m$^2_{21}$ better.
    \item Oscillation tomography of the Earth will need solar neutrinos to be studied to get a better picture of oscillated neutrino flux.
    \item Minimization of uncertainties in the fiducial volume due to the vertex shift and uncertainty in energy scale due to water transparency in Cherenkov signals.
    \item Detection of neutrinos from $\textit{hep}$ reaction since their contribution to neutrino flux is very small.
    \item Precision in the measurements of neutrinos produced from CNO cycle.
    \item Uncertainties in solar models affect the predictions of solar neutrino fluxes.
\end{itemize}
Upcoming experiments addressing the above challenges are, SNO+ (liquid scintillator) \cite{sno,snoo+}, JUNO (linear alkylbenzene) \cite{juno}, Hyper-Kamiokande (water Cherenkov) \cite{hykam}, DUNE (liquid argon) \cite{dune}, DARWIN (liquid xenon) \cite{darwinpp}.
\begin{table}[h!]
\centering
\begin{tabular}{lc c lc lc lc}
\hline
\hline
\textbf{Experiment} & \textbf{Material}&  \textbf{Reaction}& \textbf{Threshold} & \textbf{Ref.} \\ [0.5ex] % 
&&&(MeV)&\\
\hline
\hline
SNO & D$_2$O  & $\nu_e$+d$\rightarrow$e$^-$+p+p & 3.5 & \cite{snoa}\\
&&$\nu_x$+d$\rightarrow$$\nu_x$+p+n&\\
&&$\nu_x$+e$^-$$\rightarrow$$\nu_x$+e$^-$&\\
&&&&\\
SK & H$_2$O & $\nu_x$+e$^-$$\rightarrow$$\nu_x$+e$^-$ & 3.5 & \cite{ski} \\
KamLAND & LS & & 0.5/5.5 & \cite{kamsol}\\
\hline
Homestake & C$_2$Cl$_4$  & $\nu_e$+$^{37}$Cl$\rightarrow$e$^-$+$^{37}$Ar & 0.814 & \\
Borexino & LS  & $\nu_x$+e$^-$$\rightarrow$$\nu_x$+e$^-$ & 0.19 & \cite{borexv,borex,bor}\\
SAGE & $^{71}$Ga  & $\nu_e$+$^{71}$Ga$\rightarrow$$^{71}$Ge+e$^-$ & 0.233 & \cite{sage} \\
GALLEX-GNO & GaCl$_3$  & $\nu_e$+$^{71}$Ga$\rightarrow$$^{71}$Ge+e$^-$ & 0.233 & \cite{sage} \\
\hline
\hline
\end{tabular}
\caption{Solar neutrino experiments. (LS: liquid scintillator, x = e,$\mu,\tau$).}
\label{table:table4}
\end{table}
\subsubsection{Atmospheric neutrino experiment}
\begin{figure}[h!]
    \centering
    \includegraphics[width=8cm]{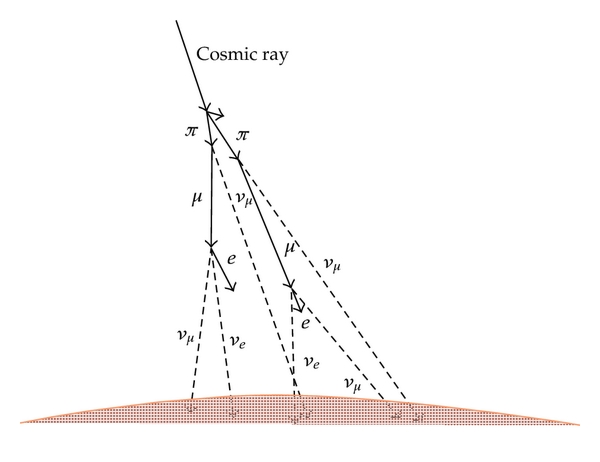}
    \hspace{0.35cm}
     \includegraphics[width=7cm]{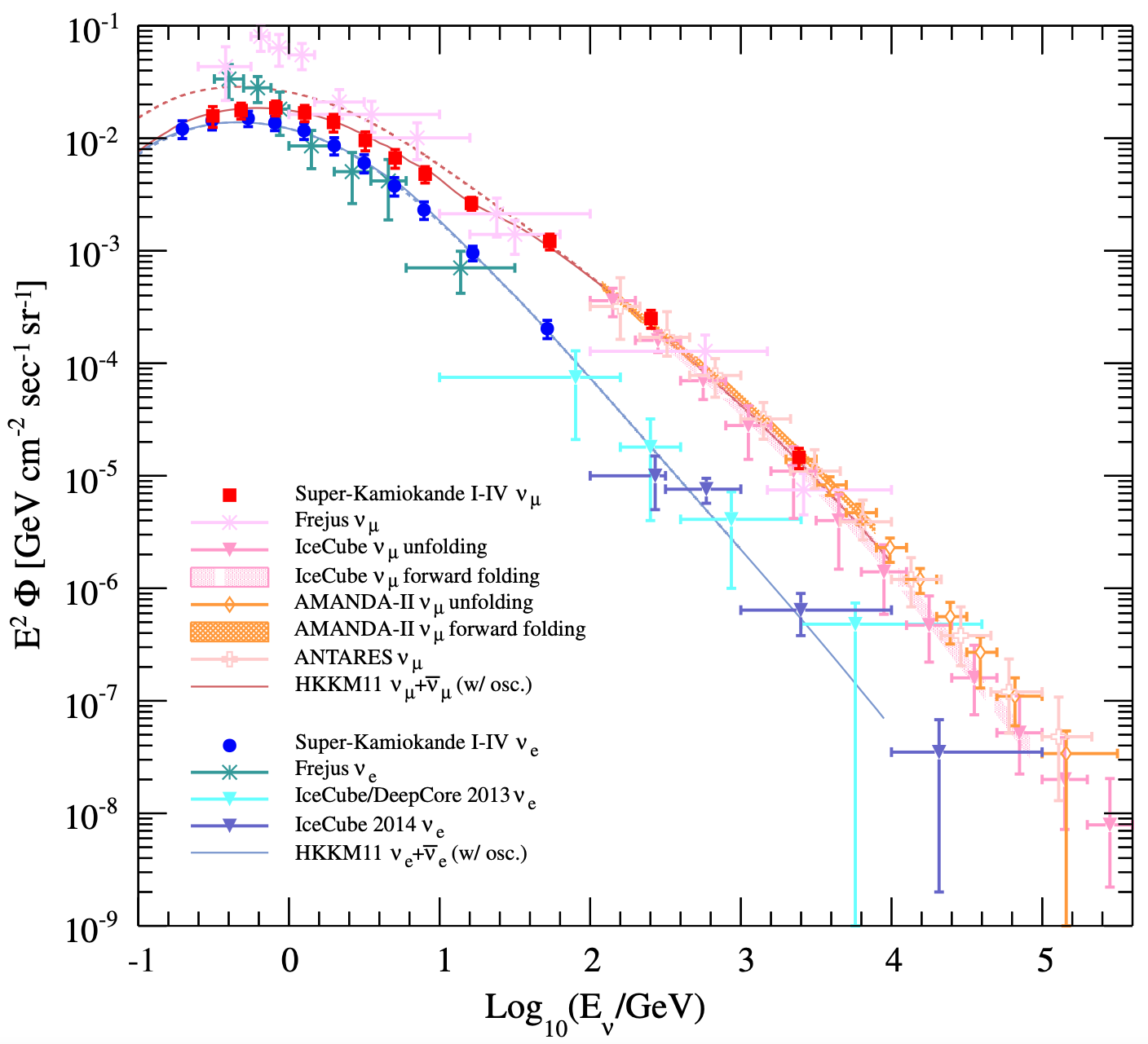}
    \caption{\textbf{Left} : production of atmospheric neutrinos. Image credit \cite{kajitas}. \textbf{Right} : Atmospheric neutrino flux. Image credit \cite{superkamnu}.}
    \label{fig:atmflux}
\end{figure}

Cosmic ray particles are mostly protons, these protons after entering the earth atmosphere interacts with atmospheric nuclei present at high altitude. These high energy nuclear interactions produce many pi mesons and less abundantly produced kaons. These meson are unstable and decay into other particles. The $\pi^+$ meson decays into a $\mu^+$ and a $\nu_{\mu}$. This produced $\mu^+$ are also unstable particles which further decays into an e$^+$, $\nu_{e}$ and $\overline{\nu_{\mu}}$ as shown in figure \ref{fig:atmflux} left. Similar decay process takes place for unstable $\pi^-$ meson and kaons. The neutrino produced in these processes are known as atmospheric neutrinos. The atmospheric flux consist of both neutrinos and antineutrinos. 
\\ Atmospheric neutrino flux as a function of neutrino energy is shown in figure \ref{fig:atmflux} right. The energy of these atmospheric neutrinos varies from few MeV to few PeV range and their path lengths are suitable to probe many of the prevailing neutrino puzzles. When these neutrinos (antineutrinos) pass via Earth the matter effects \cite{matsun} influence the oscillation probability as the oscillation parameters sin$^2$$\theta_{13}$ and $\Delta$m$^2_{32}$ are replaced by their matter equivalents. Matter effects plays a significant role in distinguishing neutrino mass hierarchy since atmospheric neutrino flux has L/E dependency. Current atmospheric neutrino experiments are listed in table $\ref{table:table5}$. 
\\ Upcoming experiments sensitive to neutrino mass ordering are, Hyper-Kamiokande (4.0 $\sigma$ for runtime of 10 years) \cite{hk2018,neu}, DUNE (3.0 $\sigma$ for runtime of 10 years) \cite{dune2020,neu}, KM3NeT/ORCA (4.4 $\sigma$ for normal ordering and 2.3 $\sigma$ for inverted ordering for runtime of 3 years) \cite{km3net2018,neu}, IceCube Upgrade (3.8 $\sigma$ for normal ordering and 1.8 $\sigma$ for inverted ordering for runtime of 6 years) \cite{ice2020,neu}.
\\
\\
\\
\textbf{Main challenges in the measurement of mass squared difference from atmospheric neutrinos}
\begin{itemize}
    \item Large fiducial mass of detector to reduce the statistical error.
    \item Underground deployment of detector to reduce the cosmic muon flux.
    \item Uncertainty in atmospheric neutrino flux since flavor changes with neutrino energy.
    \item Smearing in neutrino energy and neutrino direction measurement.
    \item Neutrino flavor identification.
    \item Reconstruction of the direction of neutrino energy.
    \item Degeneracy due to uncertainty in neutrino parameters, \cite{smirnov}.
\end{itemize}
Upcoming experiments under development addressing above challenging are, KM3NeT/ORCA (water cherenkov) \cite{kmne}, IceCube-Gen2 (ice cherenkov) \cite{iceg}, INO (iron) \cite{ino}, Hyper-Kamiokande (water cherenkov) \cite{hykam}.
\\ 
\begin{table}[h!]
\centering
\begin{tabular}{lc c lc lc}
\hline
\hline
\textbf{Experiment} & \textbf{Material}& \textbf{$\Delta$}\textbf{m}$^2_{32}$ & \textbf{Ref.} \\ [0.5ex] %
&&\scriptsize{(10$^{-3}$) eV$^2$}&\\
\hline
\hline
Super-Kamiokande & H$_2$O & 2.50$^{+0.13}_{-0.20}$ \scriptsize{(NO)} & \cite{skam}\\
&&&\\
IceCube & Ice & 2.31$^{+0.11}_{-0.13}$ \scriptsize{(NO)} & \cite{ice}\\
&&&\\
ANTARES & H$_2$O & 2.0$^{+0.4}_{-0.3}$ & \cite{anta}\\
\hline
Kamiokande & H$_2$O& & \cite{ss3} \\
Soudan2 & Fe & &\cite{ds9} \\
IMB & H$_2$O & & \cite{casp} \\
\hline
\hline
\end{tabular}
\caption{Atmospheric neutrino experiments.}
\label{table:table5}
\end{table}
\subsubsection{Reactor neutrino experiment}
Along with the energy production by nuclear fission the nuclear reactors also produces flavor pure source of anti neutrino ($\bar{\nu_e}$) flux, which are well understood and this special feature makes reactors a “free” and copious neutrino source for the study. In reactor neutrino physics we use inverse beta decay (IBD) where antineutrino will interact with the proton of detector target and produce a positron which annihilates an electron (prompt signal) and a neutron which is captured afterwards (delayed signal). 
\\ We can broadly categorize the reactor experiments into (i) short baseline ($\sim$1 km)  and (ii) long baseline ($\sim$100-1000 km) reactor experiments.  Three short baseline reactor neutrino experiments which looked for antineutrino disappearance with the main objective  to measure last unknown neutrino oscillation angle $\theta_{13}$ are Double Chooz in France \cite{dchooz}, RENO in South Korea \cite{reno1} and Daya Bay in China \cite{daya2}. All the three experiments used detectors which included liquid scintillator target loaded with 0.1$\%$ of Gadolinium. The results of the three experiments for sin$^22\theta_{13}$ are Double Chooz: 0.102 $\pm$ 0.012 \cite{dchooz20}; Daya Bay: 0.0856 $\pm$ 0.0029 \cite{daya3}; RENO: 0.0892 $\pm$ 0.0044 (stat.) $\pm$ 0.0045 (sys.) \cite{reno3}. These experiments can also add knowledge to the value of  the effective combination of mass, which can be expressed as,
\\
\\ ${}\hspace{5cm}$ $\Delta$m$^2_{ee}$ $\equiv$ cos$^2\theta_{12}$$\Delta$m$^2_{31}$ + sin$^2\theta_{12}$$\Delta$m$^2_{32}$ 
\\
\\
At the same time we can also extract information regarding the sign (+ for NO, - for IO) of a phase $\Phi_{\odot }$ which depends on solar parameters. The upcoming reactor experiment JUNO have the potential to determine the neutrino mass ordering at $\geq {3\sigma}$ to be 31$\% $ by 2030 \cite{juno2021}. Current reactor neutrino experiments shown in table $\ref{table:table6}$.
\\
\\
\textbf{Main challenges in the measurement of mass squared difference from reactor neutrinos}
\begin{itemize}
    \item Only $\overline{\nu_e}$ disappearance channel can be analyzed.
    \item Decrement of reactor neutrino flux as a function of distance because antineutrino flux is isotropic.
    \item Difficulty in computation of $\overline{\nu_e}$ spectrum, because neutrino spectrum of each decay isotope is different.
    \item About 75\% of $\overline{\nu_e}$ produced by reactor remains undetected.
    \item Suppression of neutron induced by cosmic-ray muons.
    \item Elimination of cosmogenic production of radioactive isotopes; ${}^{12}$B, ${}^{8}$Li, ${}^{6}$He inside the detector volume.
\end{itemize}
Upcoming experiments resolving above challenging are, SNO+ (${}^{130}$Te) \cite{sno}, JUNO (linear alkylbenzene) \cite{juno}.
\\
\begin{table}[h!]
\centering
\begin{tabular}{lc c lc lc}
\hline
\hline
\textbf{Experiment} & \textbf{Material} &\textbf{$\Delta$}\textbf{m}$^2_{32}$ & \textbf{Ref.} \\ [0.5ex] % 
&&\scriptsize{(10$^{-3})$ eV$^2$}&\\
\hline
\hline
Daya Bay&liquid scintillator&2.471$^{+0.068}_{-0.070}$ \scriptsize{(NO)}&\cite{daya}\\
&&-2.73$^{+0.14}_{-0.14}$ \scriptsize{(IO)}\\
\\
RENO&liquid scintillator&2.63$^{+0.14}_{-0.14}$ \scriptsize{(NO)}&\cite{reno}\\
&&-2.73$^{+0.14}_{-0.14}$ \scriptsize{(IO)}\\
\hline
Double Chooz&liquid scintillator&$\theta_{13}$&\cite{chooz}\\
KamLAND&liquid scintillator& $\theta_{12}$, $\Delta$m$^2_{21}$&\cite{kama}\\
\hline
\hline
\end{tabular}
\caption{Reactor neutrino experiments.}
\label{table:table6}
\end{table}
\subsubsection{Accelerator neutrino experiment}
The most controlled manner to neutrino production is by means of particle accelerators. The accelerators at FNAL, CERN and J-PARC boost protons at high energies and crash into heavy target, emergent debris  would primarily be the unstable pions, resulting into the beam of $\nu_{\mu}$ and $\overline{\nu_{\mu}}$ as, $\pi^{\pm}$ $\rightarrow$ $\mu^{\pm}$ + $\nu_{\mu}$($\overline{\nu_{\mu}}$). Neutrino beam are then propagates towards the detectors. For a short-baseline neutrino experiments, MicroBooNE \cite{boone}, ICARUS \cite{icarus} and SBND \cite{sbnd} receives unoscillated neutrino flux whereas oscillated neutrino flux received by long-baseline experiments, NOvA \cite{nova}, T2K \cite{t2k} and DUNE \cite{dune}. Current accelerator neutrino experiments are shown in table $\ref{table:table7}$.
\\
\\ 
\textbf{Main challenges in the measurement of mass squared difference from accelerators neutrinos}
\begin{itemize}
    \item Large fiducial mass of detector to collect high statistics of neutrino event data.
    \item Production of intense beam of neutrinos requires high power proton accelerator.
    \item Deep underground location of detector.
    \item Reduction of proton beam related background.
    \item Large distance between neutrino source and detector.
    \item Reduction of uncertainty in mixing angle ($\theta_{23}$) and determination of its octant.
    \item Elimination of neutrons (produced from cosmogenic, ${}^{238}$U/${}^{238}$Th, etc) interacting detector volume. 
    \item Neutrino energy reconstruction due to nuclear effects and nuclear properties for e.g. pion produced via neutrino interaction give rise to fake neutrino events, \cite{jyot1,jyot2,jyot3,jyot4}.
\end{itemize}
Upcoming experiments addressing above challenges are, SBND (liquid argon) \cite{sbnd}, MOMENT (water cherenkov) \cite{moment}, PROMPT (iron) \cite{prompt}, SHiP (tungsten) \cite{ship}, DsTau (tungsten) \cite{dstau}, Hyper-Kamiokande (water cherenkov) \cite{hykam}, DUNE (liquid argon) \cite{dune}.
\\
\begin{table}[h!]
\centering
\begin{tabular}{lc c lc lc}
\hline
\hline
\textbf{Experiment} & \textbf{Material} &\textbf{$\Delta$}\textbf{m}$^2_{32}$ & \textbf{Ref.} \\ [0.5ex] % 
&&\scriptsize{(10$^{-3}$) eV$^2$}&\\
\hline
\hline
NOvA&liquid scintillator&2.48$^{+0.11}_{-0.06}$ \scriptsize{(NO)}&\cite{nova}\\
&&-2.54$^{+0.11}_{-0.06}$ \scriptsize{(IO)}\\
\\
T2K&water Cherenkov&2.45$^{+0.07}_{-0.07}$ \scriptsize{(NO)}&\cite{t2k}\\
&&-2.43$^{+0.07}_{-0.07}$ \scriptsize{(IO)}\\
\\
MINOS+&steel scintillator&2.40$^{+0.08}_{-0.09}$ \scriptsize{(NO)}&\cite{minos}\\
&&-2.45$^{+0.07}_{-0.8}$ \scriptsize{(IO)}\\
\hline
ICARUS T600 & liquid argon& $\nu_{\mu}$ $\rightarrow$ $\nu_e$ & \cite{icarus} \\
SND$@$LHC & tungsten & $\nu_{\mu}$ & \cite{snd}\\
FASER$\nu$ & tungsten & $\nu_{\mu}$, $\overline{\nu_{\mu}}$ & \cite{faser} \\
\hline
\hline
\end{tabular}
\caption{Accelerator neutrino experiments.}
\label{table:table7}
\end{table}
\section{Conclusion}
To summarize, determination of neutrino mass is a difficult task. We have given a brief overview of current experimental challenges in a neutrino mass measurement. Current limits on effective electron anti-neutrino mass from $\beta$ decay by KATRIN and effective Majorana neutrino mass from 0$\nu\beta\beta$ decay by KamLAND-Zen, GERDA, Majorana Demonstrator, EXO-200, CUORE, CUPID-0, and CUPID-Mo are presented. Current bounds on neutrino mass-squared differences from neutrino oscillation by SNO, Super Kamiokande, KamLAND, IceCube, ANTARES, Daya Bay, RENO, NOvA, T2K, and MINOS+ are discussed. Present bounds on sum of neutrino masses from cosmological measurements by Planck combined with BAO, RSD, Pantheon, DES, Lyman $\alpha$, are discussed. 
\\ Given the effort of many experiments, a measurement of the absolute neutrino mass may be around the corner, especially considering cosmology. And given the interplay of all the observables the underlying model can be tested.
\\
\section*{Acknowledgements}
The authors received no financial supports or grants for this article. \\
This article is submitted to arXiv with the arXiv ID [2305.12654 [hep-ex]].
\section*{Data Availability Statement}
No data is generated in this article.\\
\section*{Copyright}
This manuscript is funded by SCOAP$^3$. \\
\bibliographystyle{unsrtnat}

\bibliography{references}

\begin{thebibliography}{141}
\providecommand{\natexlab}[1]{#1}
\providecommand{\url}[1]{\texttt{#1}}
\expandafter\ifx\csname urlstyle\endcsname\relax
  \providecommand{\doi}[1]{doi: #1}\else
  \providecommand{\doi}{doi: \begingroup \urlstyle{rm}\Url}\fi

\bibitem[Pauli()]{pauli}
W.~Pauli.
\newblock {"Brief an die Gruppe der Radioaktiven ....", Zurich 4 December
  1930}.

\bibitem[REINES and COWANjun.()]{REINES:1956ug}
F.~REINES and C.~L. COWANjun.
\newblock {\href{https://doi.org/10.1038/178446a0}{Nature $\textbf{178}$,
  446--449 (1956)}}.

\bibitem[Reines and Cowan()]{PhysRev.92.830}
F.~Reines and C.~L. Cowan.
\newblock {\href{https://link.aps.org/doi/10.1103/PhysRev.92.830}{Phys. Rev.
  $\textbf{92}$ 830--831 (1953)}}.

\bibitem[Glashow()]{GLASHOW1961579}
S.~L. Glashow.
\newblock {\href{https://doi.org/10.1016/0029-5582(61)90469-2}{Nucl. Phys.
  $\textbf{22}$, 579-588 (1961)}}.

\bibitem[Weinberg()]{PhysRevLett.19.1264}
S.~Weinberg.
\newblock {\href{https://doi.org/10.1103/PhysRevLett.19.1264}{Phys. Rev. Lett.
  $\textbf{19}$, 1264--1266 (1967)}}.

\bibitem[Majorana()]{Majorana:2008vb}
E.~Majorana.
\newblock {\href{https://doi.org/10.1007/BF02961314}{Il Nuovo Cimento
  (1924-1942) $\textbf{14}$, 171 (1937)}}.

\bibitem[Furry()]{PhysRev.56.1184}
W.~H. Furry.
\newblock {\href{https://link.aps.org/doi/10.1103/PhysRev.56.1184}{Phys. Rev.
  $\textbf{56}$, 1184--1193 (1939)}}.

\bibitem[Bilenky()]{Bilenky:2018hbz}
S.~M. Bilenky.
\newblock \emph{{Introduction to the Physics of Massive and Mixed Neutrinos,
  Vol. 947 Springer, 2018}}.

\bibitem[Fukuda et~al.({\natexlab{a}})]{PhysRevLett.81.1562}
Y.~Fukuda et~al.
\newblock {(Super-Kamiokande),
  \href{https://doi.org/10.1103/PhysRevLett.81.1562}{Phys. Rev. Lett.
  $\textbf{81}$, 1562 (1998)}}.
\newblock {\natexlab{a}}.

\bibitem[Ahmad et~al.()]{PhysRevLett.89.011301}
Q.~R. Ahmad et~al.
\newblock {(SNO),
  \href{https://link.aps.org/doi/10.1103/PhysRevLett.89.011301}{Phys. Rev.
  Lett. $\textbf{89}$, 011301 (2002)}}.

\bibitem[Kajita({\natexlab{a}})]{RevModPhys.88.030501}
T.~Kajita.
\newblock {\href{https://link.aps.org/doi/10.1103/RevModPhys.88.030501}{Rev.
  Mod. Phys. $\textbf{88}$, 030501 (2016)}}.
\newblock {\natexlab{a}}.

\bibitem[McDonald()]{RevModPhys.88.030502}
A.~B. McDonald.
\newblock {\href{https://link.aps.org/doi/10.1103/RevModPhys.88.030502}{Rev.
  Mod. Phys. $\textbf{88}$, 030502 (2016)}}.

\bibitem[J.~Froustey and Volpe()]{mv16}
C.~Pitrou J.~Froustey and M.~C. Volpe.
\newblock {\href{https://doi.org/10.1088/1475-7516/2020/12/015}{JCAP
  $\textbf{12}$, 015 (2020)}}.

\bibitem[Bennett et~al.()]{mv17}
J.~J. Bennett et~al.
\newblock {\href{https://doi.org/10.1088/1475-7516/2021/04/073}{JCAP
  $\textbf{04}$, 073 (2021)}}.

\bibitem[Akita and Yamaguchi({\natexlab{a}})]{kensu}
K.~Akita and M.~Yamaguchi.
\newblock {\href{https://dx.doi.org/10.1088/1475-7516/2020/08/012}{JCAP
  $\textbf{08}$, 012 (2020)}}.
\newblock {\natexlab{a}}.

\bibitem[Akita and Yamaguchi({\natexlab{b}})]{kensu2}
K.~Akita and M.~Yamaguchi.
\newblock {\href{https://doi.org/10.3390/universe8110552}{Universe
  $\textbf{8(11)}$, 552 (2022)}}.
\newblock {\natexlab{b}}.

\bibitem[Workman et~al.()]{ParticleDataGroup:2020ssz}
R.L. Workman et~al.
\newblock {(Particle Data Group), \href{https://pdg.lbl.gov}{Prog. Theor. Exp.
  Phys. $\textbf{2022}$, 083C01 (2022) and 2023 update}}.

\bibitem[Lesgourgues et~al.()]{julien}
J.~Lesgourgues et~al.
\newblock {\href{https://doi.org/10.1017/CBO9781139012874}{Neutrino Cosmology,
  Cambridge University Press, (2013)}}.

\bibitem[Lesgourgues and Pastor({\natexlab{a}})]{pastor}
J.~Lesgourgues and S.~Pastor.
\newblock {\href{https://doi.org/10.1016/j.physrep.2006.04.001}{Phys. Rept.
  $\textbf{429}$ 307-379 (2006)}}.
\newblock {\natexlab{a}}.

\bibitem[Lattanzi and Gerbino()]{gerbino}
M.~Lattanzi and M.~Gerbino.
\newblock {\href{https://doi.org/10.3389/fphy.2017.00070}{Front. in Phys.
  $\textbf{5}$ 70 (2018)}}.

\bibitem[Gerbino et~al.()]{synergy}
M.~Gerbino et~al.
\newblock {\href{https://doi.org/10.48550/arXiv.2203.07377}{arXiv:2203.07377v2
  [hep-ph] (2022)}}.

\bibitem[Aghamousa et~al.()]{mnj}
A.~Aghamousa et~al.
\newblock {(DESI), \href{https://arxiv.org/abs/1611.00036}{arXiv:1611.00036
  (2016)}}.

\bibitem[Laureijs et~al.()]{bhj}
R.~Laureijs et~al.
\newblock {(EUCLID), \href{https://arxiv.org/abs/1110.3193}{arXiv:1110.3193
  (2011)}}.

\bibitem[Abell et~al.()]{azxc}
P.~A. Abell et~al.
\newblock {(LSST), \href{https://arxiv.org/abs/0912.0201}{arXiv:0912.0201
  (2009)}}.

\bibitem[{J. Bock and SPHEREx Science Team}()]{klj}
{J. Bock and SPHEREx Science Team}.
\newblock {American Astronomical Society Meeting Abstracts 231, volume 231 of
  American Astronomical Society Meeting Abstracts, 354.21 (2018)}.

\bibitem[mnh()]{mnhu}
{\href{http://www.skatelescope.org}{http://www.skatelescope.org}}.

\bibitem[Ade et~al.()]{sde}
Peter Ade et~al.
\newblock {(Simons Observatory),
  \href{https://doi.org/10.1088/1475-7516/2019/02/056}{JCAP $\textbf{02}$ 056
  (2019)}}.

\bibitem[Abazajian et~al.()]{fgfs}
K.~N. Abazajian et~al.
\newblock {(CMB-S4), \href{https://arxiv.org/abs/1610.02743}{arXiv:1610.02743
  (2016)}}.

\bibitem[Hazumi et~al.()]{dfw}
M.~Hazumi et~al.
\newblock {\href{https://doi.org/10.1007/s10909-019-02150-5 }{J. Low Temp.
  Phys. $\textbf{194}$, 5-6, 443 (2019)}}.

\bibitem[Aghanim et~al.()]{cosmu1}
N.~Aghanim et~al.
\newblock {(Planck), \href{https://doi.org/10.1051/0004-6361/201833910}{A\&A
  $\textbf{641}$, A6 (2020)}}.

\bibitem[Alam et~al.()]{des}
S.~Alam et~al.
\newblock {(eBOSS), \href{https://doi.org/10.1103/PhysRevD.103.083533}{Phys.
  Rev. D $\textbf{103}$, 083533 (2021)}}.

\bibitem[Palanque-Delabrouille et~al.()]{hsa}
N.~Palanque-Delabrouille et~al.
\newblock {\href{https://doi.org/10.1088/1475-7516/2020/04/038}{JCAP
  $\textbf{04}$, 038 (2020)}}.

\bibitem[Abbott et~al.()]{dsds}
T.~M.~C. Abbott et~al.
\newblock {(DES), \href{https://doi.org/10.1103/PhysRevD.105.023520}{Phys. Rev.
  D $\textbf{105}$, 023520 (2022)}}.

\bibitem[Lesgourgues and Pastor({\natexlab{b}})]{aqqw}
J.~Lesgourgues and S.~Pastor.
\newblock {\href{https://doi.org/10.1155/2012/608515}{Advances in High Energy
  Physics $\textbf{2012}$, 608515 (2012)}}.
\newblock {\natexlab{b}}.

\bibitem[Petcov()]{cxd}
S.~T. Petcov.
\newblock {\href{https://doi.org/10.1155/2013/852987}{Advances in High Energy
  Physics $\textbf{2013}$, 852987 (2013)}}.

\bibitem[Mohapatra()]{ghg}
R.~N. Mohapatra.
\newblock {\href{https://doi.org/10.1016/S0920-5632(00)00957-9}{Nuclear Physics
  B - Proceedings Supplements $\textbf{91}$, 313-320 (2001)}}.

\bibitem[Services()]{nndc}
Nuclear~Data Services.
\newblock {International Atomic Energy Agency, https://www-nds.iaea.org/}.

\bibitem[on~behalf of~the KATRIN~collaboration()]{sh3h}
A.~Onillon on~behalf of~the KATRIN~collaboration.
\newblock {(KATRIN), \href{https://doi.org/10.1051/epjconf/202328201011}{EPJ
  Web of Conferences \textbf{282}, 01011 (2023)}}.

\bibitem[Aker et~al.({\natexlab{a}})]{katrin}
M.~Aker et~al.
\newblock {(KATRIN), \href{https://doi.org/10.1038/s41567-021-01463-1}{Nat.
  Phys. $\textbf{18}$, 160--166 (2022)}}.
\newblock {\natexlab{a}}.

\bibitem[Robertson and Knapp()]{beta8}
R.~G.~H. Robertson and D.~A. Knapp.
\newblock
  {\href{https://www.annualreviews.org/doi/10.1146/annurev.ns.38.120188.001153}{Ann.
  Rev. Nucl. Part. Sci. $\textbf{38}$, 185-215 (1988)}}.

\bibitem[Betti et~al.()]{Betti_2019}
M.G. Betti et~al.
\newblock {\href{https://doi.org/10.1088/1475-7516/2019/07/047}{JCAP
  $\textbf{07}$, 047 (2019)}}.

\bibitem[Monreal and Formaggio()]{PhysRevD.80.051301}
B.~Monreal and J.~A. Formaggio.
\newblock {\href{https://link.aps.org/doi/10.1103/PhysRevD.80.051301}{Phys.
  Rev. D $\textbf{80}$, 051301 (2009)}}.

\bibitem[Gastaldo et~al.()]{Gastaldo:2014un}
L.~Gastaldo et~al.
\newblock {\href{https://doi.org/10.1007/s10909-014-1187-4}{Journal of Low
  Temperature Physics $\textbf{176}$, 876--884 (2014)}}.

\bibitem[Alpert et~al.()]{Alpert:2015vi}
B.~Alpert et~al.
\newblock {\href{https://doi.org/10.1140/epjc/s10052-015-3329-5}{Eur. Phys. J.
  $\textbf{C75}$ 112 (2015)}}.

\bibitem[Croce et~al.()]{Croce:2016uq}
M.~P. Croce et~al.
\newblock {\href{https://doi.org/10.1007/s10909-015-1451-2}{Journal of Low
  Temperature Physics $\textbf{184}$, 958--968 (2016)}}.

\bibitem[Aker et~al.({\natexlab{b}})]{kat1}
M.~Aker et~al.
\newblock {(KATRIN),
  \href{https://doi.org/10.1103/PhysRevLett.123.221802}{Phys. Rev. Lett.
  $\textbf{123}$, 221802 (2019)}}.
\newblock {\natexlab{b}}.

\bibitem[Schechter and Valle()]{valle}
J.~Schechter and J.~W.~F. Valle.
\newblock {\href{https://doi.org/10.1103/PhysRevD.25.2951}{Phys. Rev. D
  $\textbf{25}$, 2951 (1982)}}.

\bibitem[Agostini et~al.({\natexlab{a}})]{sci}
M.~Agostini et~al.
\newblock {(GERDA), \href{https://doi.org/10.1126/science.aav8613}{Science
  $\textbf{365}$, 1445 (2019)}}.
\newblock {\natexlab{a}}.

\bibitem[Agostini et~al.({\natexlab{b}})]{PhysRevLett.125.252502}
M.~Agostini et~al.
\newblock {(GERDA),
  \href{https://link.aps.org/doi/10.1103/PhysRevLett.125.252502}{Phys. Rev.
  Lett. $\textbf{125}$, 252502 (2020)}}.
\newblock {\natexlab{b}}.

\bibitem[Azzolini et~al.()]{bn}
O.~Azzolini et~al.
\newblock {(CUPID-0),
  \href{https://doi.org/10.1103/PhysRevLett.129.111801}{Phys. Rev. Lett.
  $\textbf{129}$, 111801 (2022)}}.

\bibitem[Armengaud et~al.()]{PhysRevLett.126.181802}
E.~Armengaud et~al.
\newblock {(CUPID-Mo),
  \href{https://link.aps.org/doi/10.1103/PhysRevLett.126.181802}{Phys. Rev.
  Lett. $\textbf{126}$, 181802 (2021)}}.

\bibitem[Adams et~al.({\natexlab{a}})]{bgh}
D.~Q. Adams et~al.
\newblock {(CUORE), \href{https://doi.org/10.1038/s41586-022-04497-4}{Nature
  $\textbf{604}$, 53–58 (2022)}}.
\newblock {\natexlab{a}}.

\bibitem[Gando et~al.({\natexlab{a}})]{PhysRevLett.117.082503}
A.~Gando et~al.
\newblock {(KamLAND-Zen),
  \href{https://link.aps.org/doi/10.1103/PhysRevLett.117.082503}{Phys. Rev.
  Lett. $\textbf{117}$, 082503 (2016)}}.
\newblock {\natexlab{a}}.

\bibitem[Anton et~al.()]{PhysRevLett.123.161802}
G.~Anton et~al.
\newblock {(EXO-200),
  \href{https://doi.org/10.1103/PhysRevLett.123.161802}{Phys. Rev. Lett.
  $\textbf{123}$, 161802 (2019)}}.

\bibitem[Umehara et~al.()]{heer}
S~Umehara et~al.
\newblock {\href{https://doi.org/10.1088/1742-6596/120/5/052058}{J. Phys.:
  Conf. Ser. $\textbf{120}$ 052058 (2008)}}.

\bibitem[Argyriades et~al.()]{ssdf}
J.~Argyriades et~al.
\newblock {(NEMO-3),
  \href{https://doi.org/10.1016/j.nuclphysa.2010.07.009}{Nucl. Phys. A
  $\textbf{847}$ 168 (2010)}}.

\bibitem[Barabash et~al.()]{serr}
A.~S. Barabash et~al.
\newblock {\href{https://doi.org/10.1103/PhysRevD.98.092007}{Phys. Rev. D
  $\textbf{98}$, 092007 (2018)}}.

\bibitem[Arnold et~al.()]{fdf}
R.~Arnold et~al.
\newblock {(NEMO-3), \href{https://doi.org/10.1103/PhysRevD.94.072003}{Phys.
  Rev. D $\textbf{94}$, 072003 (2016)}}.

\bibitem[Abe et~al.({\natexlab{a}})]{zen}
S.~Abe et~al.
\newblock {(KamLAND-Zen),
  \href{https://doi.org/10.1103/PhysRevLett.130.051801}{Phys. Rev. Lett.
  $\textbf{130}$, 051801 (2023)}}.
\newblock {\natexlab{a}}.

\bibitem[Chambers et~al.()]{chamber}
C.~Chambers et~al.
\newblock {(nEXO), \href{https://doi.org/10.1038/s41586-019-1169-4}{Nature
  $\textbf{569}$, 203–207 (2019)}}.

\bibitem[Poda and Giuliani()]{denys}
D.~Poda and A.~Giuliani.
\newblock {\href{https://doi.org/10.1142/S0217751X17430126}{Int. J. Mod. Phys.
  A $\textbf{32}$, 30, 1743012 (2017)}}.

\bibitem[Mirza()]{mirza}
M.~Ibrahim Mirza.
\newblock {(LEGEND),
  \href{https://doi.org/10.48550/arXiv.2209.07598}{arXiv:2209.07598 (2022)}}.

\bibitem[Abgrall et~al.()]{LEGEND:2021bnm}
N.~Abgrall et~al.
\newblock {(LEGEND), \href{https://arxiv.org/abs/2107.11462}{arXiv: 2107.11462
  (2021)}}.

\bibitem[Oh()]{amore}
Y.~Oh.
\newblock {(AMoRE), \href{https://doi.org/10.1088/1742-6596/2156/1/012146}{J.
  Phys.: Conf. Ser. $\textbf{2156}$ 012146 (2021)}}.

\bibitem[Adhikari et~al.()]{Adhikari_2021}
G~Adhikari et~al.
\newblock {(nEXO), \href{https://doi.org/10.1088/1361-6471/ac3631}{J. Phys. G:
  Nucl. Part. Phys. $\textbf{49}$, 015104 (2021)}}.

\bibitem[Albanese et~al.({\natexlab{a}})]{sno}
V.~Albanese et~al.
\newblock {(SNO+), \href{https://doi.org/10.1088/1748-0221/16/08/P08059}{JINST
  $\textbf{16}$, P08059 (2021)}}.
\newblock {\natexlab{a}}.

\bibitem[Jeremie()]{supernemo}
A.~Jeremie.
\newblock
  {\href{https://doi.org/10.1016/j.nima.2019.04.069}{Nucl.Instrum.Meth.A
  $\textbf{958}$ 162115 (2020)}}.

\bibitem[Gando et~al.({\natexlab{b}})]{kam}
Y.~Gando et~al.
\newblock {(KamLAND-Zen),
  \href{https://doi.org/10.1088/1748-0221/16/08/P08023}{JINST $\textbf{16}$
  P08023 (2021)}}.
\newblock {\natexlab{b}}.

\bibitem[Adams et~al.({\natexlab{b}})]{next}
C.~Adams et~al.
\newblock {(NEXT), \href{https://doi.org/10.1007/JHEP08(2021)164}{J. High
  Energ. Phys. $\textbf{2021}$, 164 (2021)}}.
\newblock {\natexlab{b}}.

\bibitem[Arnquist et~al.()]{mjd23}
I.~J. Arnquist et~al.
\newblock {(MAJORANA),
  \href{https://doi.org/10.1103/PhysRevLett.130.062501}{Phys. Rev. Lett.
  $\textbf{130}$, 062501 (2023)}}.

\bibitem[Pontecorvo({\natexlab{a}})]{1957}
B.~Pontecorvo.
\newblock {Sov. Phys. JETP $\textbf{6}$ 429 (1957), Zh. Eksp. Teor. Fiz.
  $\textbf{33}$, 549 (1957)}.
\newblock {\natexlab{a}}.

\bibitem[Pontecorvo({\natexlab{b}})]{1958}
B.~Pontecorvo.
\newblock {Sov. Phys. JETP $\textbf{7}$, 172 (1958); Zh. Eksp. Teor. Fiz.
  $\textbf{34}$, 247 (1958)}.
\newblock {\natexlab{b}}.

\bibitem[Eguchi et~al.()]{kams}
K.~Eguchi et~al.
\newblock {(KamLAND),
  \href{https://doi.org/10.1103/PhysRevLett.90.021802}{Phys. Rev. Lett.
  $\textbf{90}$, 021802 (2003)}}.

\bibitem[Araki et~al.()]{gd}
T.~Araki et~al.
\newblock {(KamLAND),
  \href{https://doi.org/10.1103/PhysRevLett.94.081801}{Phys. Rev. Lett.
  $\textbf{94}$, 081801 (2005)}}.

\bibitem[Acero et~al.()]{nova}
M.A. Acero et~al.
\newblock {(NOvA), \href{https://doi.org/10.1103/PhysRevLett.123.151803}{Phys.
  Rev. Lett. $\textbf{123}$, 151803 (2019)}}.

\bibitem[Bilenky and Pontecorvo({\natexlab{a}})]{Bilenky:1978nj}
S.~M. Bilenky and B.~Pontecorvo.
\newblock {\href{https://doi.org/10.1016/0370-1573(78)90095-9}{Phys. Rept.
  $\textbf{41}$, 225--261 (1978)}}.
\newblock {\natexlab{a}}.

\bibitem[Bilenky and Pontecorvo({\natexlab{b}})]{Bilenky:1976wv}
S.~M. Bilenky and B.~Pontecorvo.
\newblock {\href{https://doi.org/10.1007/BF02746567}{Lettere al Nuovo Cimento
  (1971-1985) $\textbf{17}$, 569--574 (1976)}}.
\newblock {\natexlab{b}}.

\bibitem[de~Salas et~al.()]{glodata}
P.F. de~Salas et~al.
\newblock {\href{https://doi.org/10.1007/JHEP02(2021)071}{JHEP $\textbf{02}$,
  071 (2021)}}.

\bibitem[Esteban et~al.()]{sep9}
I.~Esteban et~al.
\newblock {\href{https://doi.org/10.1007/JHEP09(2020)178}{J. High Energ. Phys.
  $\textbf{2020}$, 178 (2020)}}.

\bibitem[Capozzi et~al.()]{sep90}
F.~Capozzi et~al.
\newblock {\href{https://doi.org/10.1103/PhysRevD.95.096014}{Phys. Rev. D
  $\textbf{95}$, 096014 (2017)}}.

\bibitem[Vinyoles et~al.()]{ssmb16}
Núria Vinyoles et~al.
\newblock {\href{https://doi.org/10.3847/1538-4357/835/2/202}{ApJ
  $\textbf{835}$ 202 (2017)}}.

\bibitem[Gando et~al.({\natexlab{c}})]{kama}
A.~Gando et~al.
\newblock {(KamLAND), \href{https://doi.org/10.1103/PhysRevD.88.033001}{Phys.
  Rev. D $\textbf{88}$, 033001 (2013)}}.
\newblock {\natexlab{c}}.

\bibitem[Bellerive()]{bchall1}
A.~Bellerive.
\newblock
  {\href{https://conferences.fnal.gov/lp2003/program/papers/bellerive.pdf}{Review
  of Solar Neutrino Experiments}}.

\bibitem[Bahcall()]{bchall2}
J.~N. Bahcall.
\newblock
  {\href{http://www.sns.ias.edu/~jnb/}{https://www.sns.ias.edu/$\sim$jnb/}}.

\bibitem[Appel et~al.()]{bors2}
S.~Appel et~al.
\newblock {(Borexino),
  \href{https://doi.org/10.1103/PhysRevLett.129.252701}{Phys. Rev. Lett.
  $\textbf{129}$, 252701 (2022)}}.

\bibitem[Anderson et~al.()]{snoo+}
M.~Anderson et~al.
\newblock {(SNO+), \href{https://doi.org/10.1103/PhysRevD.99.012012}{Phys. Rev.
  D $\textbf{99}$, 012012 (2019)}}.

\bibitem[Abusleme et~al.()]{juno}
A.~Abusleme et~al.
\newblock {(JUNO), \href{https://doi.org/10.1016/j.ppnp.2021.103927}{Prog.
  Part. Nucl. Phys. $\textbf{123}$, 103927 (2022)}}.

\bibitem[Kudenko()]{hykam}
Y.~Kudenko.
\newblock {\href{https://doi.org/10.1088/1748-0221/15/07/C07029}{ JINST
  $\textbf{15}$ C07029 (2020)}}.

\bibitem[Abi et~al.({\natexlab{a}})]{dune}
B.~Abi et~al.
\newblock {(DUNE), \href{https://arxiv.org/abs/2002.03005}{arXiv:2002.03005
  (2020)}}.
\newblock {\natexlab{a}}.

\bibitem[Aalbers et~al.()]{darwinpp}
J.~Aalbers et~al.
\newblock {(DARWIN),
  \href{https://doi.org/10.1140/epjc/s10052-020-08602-7}{Eur. Phys. J. C
  $\textbf{80}$, 1133 (2020)}}.

\bibitem[Athar et~al.()]{neu}
M.~S. Athar et~al.
\newblock {\href{https://doi.org/10.1016/j.ppnp.2022.103947}{Prog. Part. Nucl.
  Phys. $\textbf{124}$, 103947 (2022)}}.

\bibitem[Nakajima()]{nu20}
Y.~Nakajima.
\newblock {\href{http://dx.doi.org/10.5281/zenodo.3959639}{Talk At Neutrino,
  2020 (2020)}}.

\bibitem[Aharmim et~al.()]{snoa}
B.~Aharmim et~al.
\newblock {(SNO), \href{https://doi.org/10.1103/PhysRevC.88.025501}{Phys. Rev.
  C $\textbf{88}$, 025501 (2013)}}.

\bibitem[Abe et~al.({\natexlab{b}})]{ski}
K.~Abe et~al.
\newblock {(Super-Kamiokande),
  \href{https://doi.org/10.1103/PhysRevD.94.052010}{Phys. Rev. D $\textbf{94}$,
  052010 (2016)}}.
\newblock {\natexlab{b}}.

\bibitem[Gando et~al.({\natexlab{d}})]{kamsol}
A.~Gando et~al.
\newblock {(KamLAND), \href{http://dx.doi.org/10.1103/PhysRevC.92.055808}{Phys.
  Rev. C $\textbf{92}$, 055808 (2015)}}.
\newblock {\natexlab{d}}.

\bibitem[Agostini et~al.({\natexlab{c}})]{borexv}
M.~Agostini et~al.
\newblock {(Borexino), \href{https://doi.org/10.1038/s41586-018-0624-y}{Nature
  $\textbf{562}$, 505--510 (2018)}}.
\newblock {\natexlab{c}}.

\bibitem[Agostini et~al.({\natexlab{d}})]{borex}
M.~Agostini et~al.
\newblock {(Borexino), \href{https://doi.org/10.1103/PhysRevD.101.062001}{Phys.
  Rev. D $\textbf{101}$, 062001 (2020)}}.
\newblock {\natexlab{d}}.

\bibitem[Agostini et~al.({\natexlab{e}})]{bor}
M.~Agostini et~al.
\newblock {(Borexino), \href{https://doi.org/10.1038/s41586-020-2934-0}{Nature
  $\textbf{587}$ 577-582 (2020)}}.
\newblock {\natexlab{e}}.

\bibitem[Abdurashitov et~al.()]{sage}
J.~N. Abdurashitov et~al.
\newblock {(SAGE), \href{https://doi.org/10.1103/PhysRevC.80.015807}{Phys. Rev.
  C $\textbf{80}$, 015807 (2009)}}.

\bibitem[Blennow and Smirnov()]{matsun}
M.~Blennow and A.~Yu. Smirnov.
\newblock {\href{https://doi.org/10.1155/2013/972485}{Adv. High Energy Phys.
  $\textbf{2013}$ 972485 (2013)}}.

\bibitem[Abe et~al.({\natexlab{c}})]{hk2018}
K.~Abe et~al.
\newblock {(Hyper-Kamiokande),
  \href{https://doi.org/10.48550/arXiv.1805.04163}{arXiv:1805.04163v2
  [physics.ins-det] (2018)}}.
\newblock {\natexlab{c}}.

\bibitem[Abi et~al.({\natexlab{b}})]{dune2020}
B.~Abi et~al.
\newblock {(DUNE),
  \href{https://doi.org/10.48550/arXiv.2002.03005}{arXiv:2002.03005v2 [hep-ex]
  (2020)}}.
\newblock {\natexlab{b}}.

\bibitem[Aiello et~al.({\natexlab{a}})]{km3net2018}
S.~Aiello et~al.
\newblock {(KM3NeT),
  \href{https://doi.org/10.1140/epjc/s10052-021-09893-0}{Eur. Phys. J. C
  $\textbf{82}$, 26 (2022)}}.
\newblock {\natexlab{a}}.

\bibitem[Aartsen et~al.({\natexlab{a}})]{ice2020}
M.~G. Aartsen et~al.
\newblock {(IceCube-Gen2, JUNO),
  \href{https://doi.org/10.1103/PhysRevD.101.032006}{Phys. Rev. D
  $\textbf{101}$, 032006 (2020)}}.
\newblock {\natexlab{a}}.

\bibitem[Kajita({\natexlab{b}})]{kajitas}
T.~Kajita.
\newblock {\href{https://doi.org/10.1155/2012/504715}{Adv. High Energy Phys
  $\textbf{2012}$, 504715 (2012)}}.
\newblock {\natexlab{b}}.

\bibitem[Richard et~al.()]{superkamnu}
E.~Richard et~al.
\newblock {(Super-Kamiokande),
  \href{http://dx.doi.org/10.1103/PhysRevD.94.052001}{Phys. Rev. D
  $\textbf{94}$, 052001 (2016)}}.

\bibitem[Smirnov()]{smirnov}
A.~Smirnov.
\newblock {\href{https://doi.org/10.22323/1.196.0027}{PoS \textbf{Neutel2013},
  027 (2014)}}.

\bibitem[Aiello et~al.({\natexlab{b}})]{kmne}
S.~Aiello et~al.
\newblock {(KM3NeT),
  \href{https://doi.org/10.1140/epjc/s10052-021-09893-0}{Eur. Phys. J. C
  $\textbf{82}$, 26 (2022)}}.
\newblock {\natexlab{b}}.

\bibitem[Aartsen et~al.({\natexlab{b}})]{iceg}
M.~G. Aartsen et~al.
\newblock {(IceCube-Gen2),
  \href{https://doi.org/10.1103/PhysRevD.101.032006}{Phys. Rev. D
  $\textbf{101}$, 032006 (2020)}}.
\newblock {\natexlab{b}}.

\bibitem[Kumar et~al.()]{ino}
A~Kumar et~al.
\newblock {(INO), \href{https://doi.org/10.1007/s12043-017-1373-4}{Pramana
  $\textbf{88}$, 5, 79 (2017)}}.

\bibitem[Abe et~al.({\natexlab{d}})]{skam}
K.~Abe et~al.
\newblock {(Super-Kamiokande),
  \href{https://doi.org/10.1103/PhysRevD.97.072001}{Phys. Rev. D $\textbf{97}$,
  072001 (2018)}}.
\newblock {\natexlab{d}}.

\bibitem[Aartsen et~al.({\natexlab{c}})]{ice}
M.~G. Aartsen et~al.
\newblock {(IceCube),
  \href{https://doi.org/10.1103/PhysRevLett.120.071801}{Phys. Rev. Lett.
  $\textbf{120}$, 071801 (2018)}}.
\newblock {\natexlab{c}}.

\bibitem[Albert et~al.()]{anta}
A.~Albert et~al.
\newblock {(ANTARES), \href{https://doi.org/10.1007/JHEP06(2019)113}{JHEP
  $\textbf{06}$ 113 (2019)}}.

\bibitem[Fukuda et~al.({\natexlab{b}})]{ss3}
Y.~Fukuda et~al.
\newblock {(Kamiokande),
  \href{https://doi.org/10.1016/0370-2693(94)91420-6}{Phys. Lett.
  $\textbf{B35}$, 237 (1994)}}.
\newblock {\natexlab{b}}.

\bibitem[Sanchez et~al.()]{ds9}
M.~C. Sanchez et~al.
\newblock {(Soudan 2), \href{https://doi.org/10.1103/PhysRevD.68.113004}{Phys.
  Rev. $\textbf{D68}$, 113004 (2003)}}.

\bibitem[Casper et~al.()]{casp}
D.~Casper et~al.
\newblock {\href{https://doi.org/10.1103/PhysRevLett.66.2561}{Phys. Rev. Lett.
  $\textbf{66}$, 2561 (1991)}}.

\bibitem[Abe et~al.({\natexlab{e}})]{dchooz}
Y.~Abe et~al.
\newblock {(Double Chooz),
  \href{https://doi.org/10.1103/PhysRevLett.108.131801}{Phys. Rev. Lett.
  $\textbf{108}$, 131801 (2012)}}.
\newblock {\natexlab{e}}.

\bibitem[Ahn et~al.()]{reno1}
J.~K. Ahn et~al.
\newblock {(RENO), \href{https://doi.org/10.1103/PhysRevLett.108.191802}{Phys.
  Rev. Lett. $\textbf{108}$, 191802 (2012)}}.

\bibitem[An et~al.()]{daya2}
F.~P. An et~al.
\newblock {(Daya Bay),
  \href{https://doi.org/10.1103/PhysRevLett.108.171803}{Phys. Rev. Lett.
  $\textbf{108}$, 171803 (2012)}}.

\bibitem[on~behalf of~the Double Chooz~Collaboration()]{dchooz20}
T.~S.~Bezerra on~behalf of~the Double Chooz~Collaboration.
\newblock {\href{http://dx.doi.org/10.5281/zenodo.3959541}{Talk At Neutrino
  2020 (2020)}}.

\bibitem[Adey et~al.()]{daya3}
D.~Adey et~al.
\newblock {(Daya Bay),
  \href{https://doi.org/10.1103/PhysRevLett.121.241805}{Phys. Rev. Lett.
  $\textbf{121}$, 241805 (2018)}}.

\bibitem[on~behalf of~the RENO~Collaboration()]{reno3}
J.~Yoo on~behalf of~the RENO~Collaboration.
\newblock {\href{https://zenodo.org/record/4123573}{Talk At Neutrino 2020
  (2020)}}.

\bibitem[Forero et~al.()]{juno2021}
D.~V. Forero et~al.
\newblock {\href{https://doi.org/10.1103/PhysRevD.104.113004}{Phys. Rev. D
  $\textbf{104}$, 113004 (2021)}}.

\bibitem[Adey()]{daya}
D.~Adey.
\newblock {(Daya Bay),
  \href{https://doi.org/10.1103/PhysRevLett.121.241805}{Phys. Rev. Lett.
  $\textbf{121}$, 241805 (2018)}}.

\bibitem[Bak et~al.()]{reno}
G.~Bak et~al.
\newblock {(RENO), \href{https://doi.org/10.1103/PhysRevLett.121.201801}{Phys.
  Rev. Lett. $\textbf{121}$, 201801 (2018)}}.

\bibitem[de~Kerret et~al.()]{chooz}
H.~de~Kerret et~al.
\newblock {(Double Chooz), \href{https://doi.org/10.1038/s41567-020-0831-y
  }{Nat. Phys. $\textbf{16}$, 558--564 (2020)}}.

\bibitem[Abratenko et~al.()]{boone}
P.~Abratenko et~al.
\newblock {(MicroBooNE),
  \href{https://doi.org/10.1103/PhysRevD.105.L051102}{Phys. Rev. D
  $\textbf{105}$, L051102 (2022)}}.

\bibitem[Ali-Mohammadzadeh et~al.()]{icarus}
B.~Ali-Mohammadzadeh et~al.
\newblock {(ICARUS),
  \href{https://doi.org/10.1088/1748-0221/15/10/T10007}{JINST $\textbf{15}$
  T10007 (2020)}}.

\bibitem[Acciarri et~al.()]{sbnd}
R.~Acciarri et~al.
\newblock {(SBND), \href{https://doi.org/10.1088/1748-0221/15/06/P06033}{JINST
  $\textbf{15}$ P06033 (2020)}}.

\bibitem[Abe et~al.({\natexlab{f}})]{t2k}
K.~Abe et~al.
\newblock {(T2K), \href{https://doi.org/10.1038/s41586-020-2177-0}{Nature
  $\textbf{580}$, 339--344 (2020)}}.
\newblock {\natexlab{f}}.

\bibitem[Nagu et~al.({\natexlab{a}})]{jyot1}
S.~Nagu et~al.
\newblock {\href{https://doi.org/10.1016/j.nuclphysb.2019.114888}{Nucl. Phys. B
  $\textbf{951}$, 114888 (2020)}}.
\newblock {\natexlab{a}}.

\bibitem[Singh et~al.()]{jyot2}
J.~Singh et~al.
\newblock {\href{https://doi.org/10.1016/j.nuclphysb.2020.115103}{Nucl. Phys. B
  $\textbf{957}$, 115103 (2020)}}.

\bibitem[Nagu et~al.({\natexlab{b}})]{jyot3}
S.~Nagu et~al.
\newblock {\href{https://doi.org/10.1155/2020/5472713}{Adv. High Energy Phys
  $\textbf{2020}$, 5472713 (2020)}}.
\newblock {\natexlab{b}}.

\bibitem[Naaz et~al.()]{jyot4}
S.~Naaz et~al.
\newblock {\href{https://doi.org/10.1016/j.nuclphysb.2018.05.018}{Nucl. Phys. B
  $\textbf{933}$, 40-52 (2018)}}.

\bibitem[Vihonen()]{moment}
S.~Vihonen.
\newblock {\href{https://arxiv.org/abs/2202.05509}{arXiv:2202.05509 (2022)}}.

\bibitem[Tang et~al.()Tang, Vihonen, and Xu]{prompt}
J.~Tang, S.~Vihonen, and Y.~Xu.
\newblock {\href{https://doi.org/10.1088/1572-9494/ac5245}{Commun. Theor. Phys.
  $\textbf{74}$ 035201 (2022)}}.

\bibitem[Pastore()]{ship}
A.~Pastore.
\newblock {(SHiP), \href{https://doi.org/10.22323/1.364.0372}{PoS EPS-HEP2019
  372 (2020)}}.

\bibitem[Aoki()]{dstau}
S.~Aoki.
\newblock {(DsTau), \href{https://doi.org/10.1007/JHEP01(2020)033}{J. High
  Energ. Phys. $\textbf{2020}$, 33 (2020)}}.

\bibitem[Adamson et~al.()]{minos}
P.~Adamson et~al.
\newblock {(MINOS+),
  \href{https://doi.org/10.1103/PhysRevLett.125.131802}{Phys. Rev. Lett.
  $\textbf{125}$, 131802 (2020)}}.

\bibitem[Albanese et~al.({\natexlab{b}})]{snd}
R.~Albanese et~al.
\newblock {(SND$@$LHC),
  \href{https://doi.org/10.1103/PhysRevLett.131.031802}{Phys. Rev. Lett.
  $\textbf{131}$, 031802 (2023)}}.
\newblock {\natexlab{b}}.

\bibitem[Abreu et~al.()]{faser}
H.~Abreu et~al.
\newblock {(FASER), \href{https://doi.org/10.1103/PhysRevLett.131.031801}{Phys.
  Rev. Lett. $\textbf{131}$, 031801 (2023)}}.

\end{thebibliography}
%\addbibresource{references} %Imports bibliography file

\end{document}